\documentclass[12pt,a4paper]{article}
\usepackage{amssymb,amsmath,graphicx,subcaption,cite}

\usepackage{epstopdf}
\usepackage[utf8]{inputenc}
\usepackage[english]{babel}

\begin{document}

%-----------------------------------------------------------------------------------
	
\begin{center}
   {\Large\bf Anisotropy driven reversal of magnetisation in Blume-Capel ferromagnet: A Monte Carlo study}
\end{center}
\vskip 1 cm
\begin{center} 
    Moumita Naskar$^1$ and Muktish Acharyya$^{2,*}$
   
   \textit{Department of Physics, Presidency University,}\\
   \textit{86/1 College Street, 
           Kolkata-700073, India} 
\vskip 0.2 cm
   
   {Email$^1$:moumita1.rs@presiuniv.ac.in}\\
   {Email$^2$:muktish.physics@presiuniv.ac.in}
\end{center}
\vspace {1.0 cm}
	
%----------------------------------------------------------------------------------
\noindent {\bf Abstract:} The two dimensional Spin-1 Blume-Capel ferromagnet is
studied by Monte Carlo simulation with Metropolis algorithm. Starting from initial
ordered spin configuration the reversal of magnetisation is investigated in presence
of a magnetic field ($h$) applied in the opposite direction. The variations of the reversal
time with the strength of single site anisotropy are investigated in details.
The exponential dependence was observed. The
systematic variations of the mean reversal time with positive and negative anisotropy
was found. The mean macroscopic reversal time was observed to be linearly dependent on
a suitably defined microscopic reversal time. The saturated magnetisation $M_f$ after
the reversal was noticed to be dependent of the strength of anisotropy $D$. An interesting scaling relation was obtained, $|M_f| \sim |h|^{\beta}f(D|h|^{-\alpha})$ with the 
scaling function of the form $f(x)= \frac{1}{1+e^{(x-a)/b}}$. The temporal evolution
of density of $S_i^z=0$ (surrounded by all $S_i^z=+1$) is observed to be exponentially
decaying. The growth of mean density of $S_i^z=-1$ has been fitted in a
function $\rho_{-1}(t) \sim \frac{1}{a+e^{(t_c-t)/c}}$. 
The characteristic
time shows $t_c \sim e^{-rD}$ and a crossover in the rate of exponential falling is 
observed at $D=1.5$. The metastable volume fraction has been found to obey the Avrami's law.

\vskip 4cm

\textbf{Keywords: Blume-Capel model, Monte Carlo simulation, Metropolis 
algorithm, Magnetic anisotropy, Magnetisation reversal}

\vskip 1cm

\noindent $^*$Corresponding author, E-mail:muktish.physics@presiuniv.ac.in	
%----------------------------------------------------------------------------------
\newpage
\noindent{\large\bf I. Introduction}
\vspace {0.5 cm}

The field driven reversal of magnetisation in the magnetic materials
is an interesting field of modern research. In the technology
\cite{techno} of
magnetic recording\cite{daniel}, this scale of time is responsible for the speed
of recording. The longevity of magnetic storage device depends on the
reversal time. The dependence of this reversal time on the temperature,
disorder and various physical parameters is a serious matter of 
investigation to the theoretician as well as the experimentalists.
These results will be useful to the technologists to prepare the 
magnetic sample in such way so that the reversal time can be tuned  
to the demand of technological world.

To study the reversal of magnetisation in ferromagnetic sample, the
simplest choice would be Ising ferromagnet. A big pool 
of literatures of such
study has been developed in last few decades. In this context, the 
phenomenological Becker-\"Doring theory\cite{becker} is much appealing. Due to its
simplicity, the reversal time (or so called nucleation time) can be
obtained as function of the temperature of the system and the magnitude
of applied magnetic field. However, the prediction of this phenomenological
theory needed simulational verification to identify the reversal 
mechanism via
the coalescence of multiple droplets or by the 
growth of single supercritical droplet. The only method to verify it is
the simple Monte Carlo simulation where the growth of droplets can be
studied as phase ordering kinetics. The rate of nucleation of
crystalline solids in solid-melt system demands to be an important
historical study\cite{grant} in this context. The idea of the decay of
metastable state of such a system opened a field of research. The 
longevity of metastable states in kinetic Ising 
ferromagnet is studied\cite{rikvold1}
to find its dependence on the applied field and system size. A large
scale Monte Carlo study is done\cite{stauffer} 
using multispin coding algorithm
to investigate the behaviors of metastable
lifetimes in different spatial dimensions. A dynamical magnetisation
reversal transition with a pulsed magnetic field was studied\cite{bkc}.
The relaxation of Ising ferromagnet after a sudden reversal of applied
magnetic field is also studied\cite{binder1}
The thermally activated magnetisation switching of small ferromagnetic
particles driven by an external magnetic field has been 
investigated and interestingly a crossover from 
coherent rotation to nucleation
for a classical anisotropic Heisenberg model, has been reported\cite{uli}.
The rates of growth and decay of the clusters of
different sizes, have been studied\cite{vehkamaki} 
as functions of external field and temperature. The rate of 
nucleation of critical
nuclei, its speed of expansion and the 
corresponding changes of free energy is related 
and is described by famous
Kolmogorov–Johnson–Mehl–Avrami (KJMA) 
law\cite{kolmogorov,johnson,avrami}.The 
reversal of magnetisation assisted by heat, in ultra thin films 
for ultra-high-density information recording 
has been investigated\cite{rikvold2}.
The nucleation time was observed \cite{ma1} to increase 
in the presence of magnetic field which is spreading over the
space in time as compared to that in static field. The behaviours of
metastable states in Ising ferromagnet in the presence of a gradient
of field\cite{ma2} and in the simultaneous presence of gradients of
fields and the temperature was investigated\cite{ma3} recently. The 
dependence of metastable lifetime on the width of quench disorder was
investigated\cite{moumita} recently in random field Ising model.

The above mentioned studies are basically done on Ising (or anisotropic
Heisenberg) model. In the case of spin-${1 \over 2}$ 
Ising model, the reversal is governed microscopically
by the spin flip. The spin would
flip from +1 to -1 due to the application of external magnetic field. 
The metastability and its lifetime is function of the temperature and
the magnitude of external magnetic field. {\it What will happen if the spin
component has become perpendicular to the applied magnetic field ?}
Precisely, how does the anisotropy take part in reversal mechanism
in the spin-1 Blume-Capel\cite{blume,capel} model ? Metastability and
nucleation in the Spin-1 Blume-Capel (BC) ferromagnet was studied and found
the different mechanism of transition\cite{cirillo}. The first order
phase transition was studied\cite{costabile}
in the spin-1 BC model by effective field
theory. 
The critical point temperature was determined\cite{silva}
in the BC model by using
Wang-Landau Monte Carlo simulation.
The dynamical phase transition 
was studied\cite{gulpinar}
in a randomly diluted
single site anisotropic BC model in presence of time varying
oscillating magnetic field. The dynamical phase transition in BC model
driven by propagating and standing magnetic field wave is 
studied\cite{ajay} recently. The metastability in the BC model with 
distributed anisotropy was studied\cite{park} 
using different dynamics. The tricritical behaviour was observed in three
component spin model\cite{fisher} by renormalization group study. The first order phase transition and the 
tricritical behaviour in the BC model was also studied\cite{kwak} recently by
Wang-Landau sampling method.
The scaling and the universality was studied\cite{fytas1} in the phase diagram of two dimensional
Blume-Capel model through Monte Carlo simulation. The universality from disorder was
also studied \cite{fytas2}recently in the random bond Blume-Capel model. 
Very recently, the Blume-Capel model (in two dimension) with randomly quenched crystal field, is studied\cite{erol2} by high precision
Monte Carlo technique to conclude that this belongs to two dimensional Ising universality class. The absence of first order transition was reported\cite{sumedha}
in randomly quenched crystal field in Blume-Capel model on a fully connected graph.
The mixed spin ($S=1$,$S=1/2$) Blume-Capel model was studied\cite{selke} by Monte Carlo
simulation and absence of tricritical point was noticed in two dimensions.
The magnetic 
properties of mixed integer and half-integer spins in a Blume-Capel model was studied
\cite{bahmad} by Monte Carlo simulation.

In the present article, we have extensively investigated the role of 
single site anisotropy to the reversal mechanism in BC model. We employed
the Monte Carlo simulation using Metropolis single spin flip algorithm.
The time dependence of the metastable volume fraction is investigated.
The saturated magnetisation, after the reversal, was found to follow an
interesting scaling behaviour. We have organised the manuscript as follows:
in the section -II we have defined the BC model and the simulation method,
the results are reported briefly in section-III and the paper ends with
concluding remarks in section-IV.

%----------------------------------------------------------------------------------
\vskip 0.5 cm
\noindent {\large\bf II. The Model and Simulation method:}
\vskip 0.5 cm

\noindent The spin-1 Blume-Capel ferromagnet is modelled by the following Hamiltonian,
\begin{equation}
     H= -J\sum_{<i,j>}S_i^z S_j^z + D\sum_{i}(S_i^z)^2 - h\sum_{i}S_i^z,
\end{equation}
where $S_i^z$ is the z-component of the spin ($S=1$) at i-th lattice site. 
$S_i^z$ can assume three values, +1, 0 and -1. The first term signifies the 
contribution to the energy due to the nearest neighbours ferromagnetic ($J>0$) exchange interaction.
  Second term is considered to model the effect of single 
site anisotropy (or magnetocrystalline anisotropy arising from the crystal field 
generated by crystal structure) with strength $D$. For simplicity, we have considered the uniform $D$. The Zeeman energy involving the 
interaction of applied magnetic field ($h$) with each spin, is being represented by the third term. 
We have considered  a two dimensional ferromagnetic square lattice of 
size $L \times L$ with periodic boundary conditions applied to both directions.

Let us briefly describe the method of Monte Carlo simulation employed here.
Initially, the system is considered to 
be in perfectly ordered state where all the spins are pointing up $S_i^z=+1$ 
$\forall$ i. A site (i-th say) has been chosen randomly. The present value of 
$S_i^z$ at that chosen site is $S_i^z(initial)$. The updated value may be any
of the three (+1, 0 and -1) possible values. The final test value of $S_i^z$ is
chosen randomly from any of these three values with equal probability. Let this
test value is labeled as $S_i^z(final)$. The probability of $S_i^z$, to assume
the final value $S_i^z(final)$ from its initial value $S_i^z(initial)$ is given
by Metropolis transition probability:
\cite{binder2, metro}, 
\begin{equation}
  P(S_i^z(initial) \to S_i^z(final)) = Min[1,{\rm exp}({- \frac{\Delta H}{k T}})],
\end{equation}
\noindent where $\Delta H$ is the change in energy (calculated from equation-1) due
to the change in the value of $S_i^z$, from $S_i^z(initial)$ to $S_i^z(final)$. $k$
is the Boltzmann constant and $T$ is the temperature of the system. The temperature
of the system is measured in the unit of $J/k$. 
For simplicity, we set $J=1$ and $k=1$ throughout the simulational study.
The acceptance of the final value
$S_i^z(final)$ is determined just by comparing a random number with the Metropolis
transition probability. The test move is accepted only when the random number
(uniformly distributed in the range [0,1]) is less than or equal to
$P(S_i^z(initial) \to S_i^z(final))$. In this way total $L^2$ number of randomly
chosen spins are updated. This is usually called random updating scheme and 
$L^2$ number of random updates constitutes one Monte Carlo Step per Spin (MCSS) 
and acts as the unit of time in the problem.

For any fixed value of the temperature ($T$) of the system and 
the applied external magnetic field ($h$), the 
magnetisation of the system is determined by
\begin{equation}
  M(t) =\frac{1}{L^2} \sum_{i}^{L^2} S_i^z  
\end{equation}
\noindent To study the reversal of the magnetisation, precisely one has to calculate
the minimum time required (in MCSS), to have the negative magnetisation, starting
from a perfectly ordered configuration.
	
%----------------------------------------------------------------------------------
\vskip 2 cm
\noindent {\large\bf III. Simulational results:} 
\vskip 0.5 cm

\begin{figure}[h!]
\begin{center}
\includegraphics[angle=-90,width=0.5\textwidth]{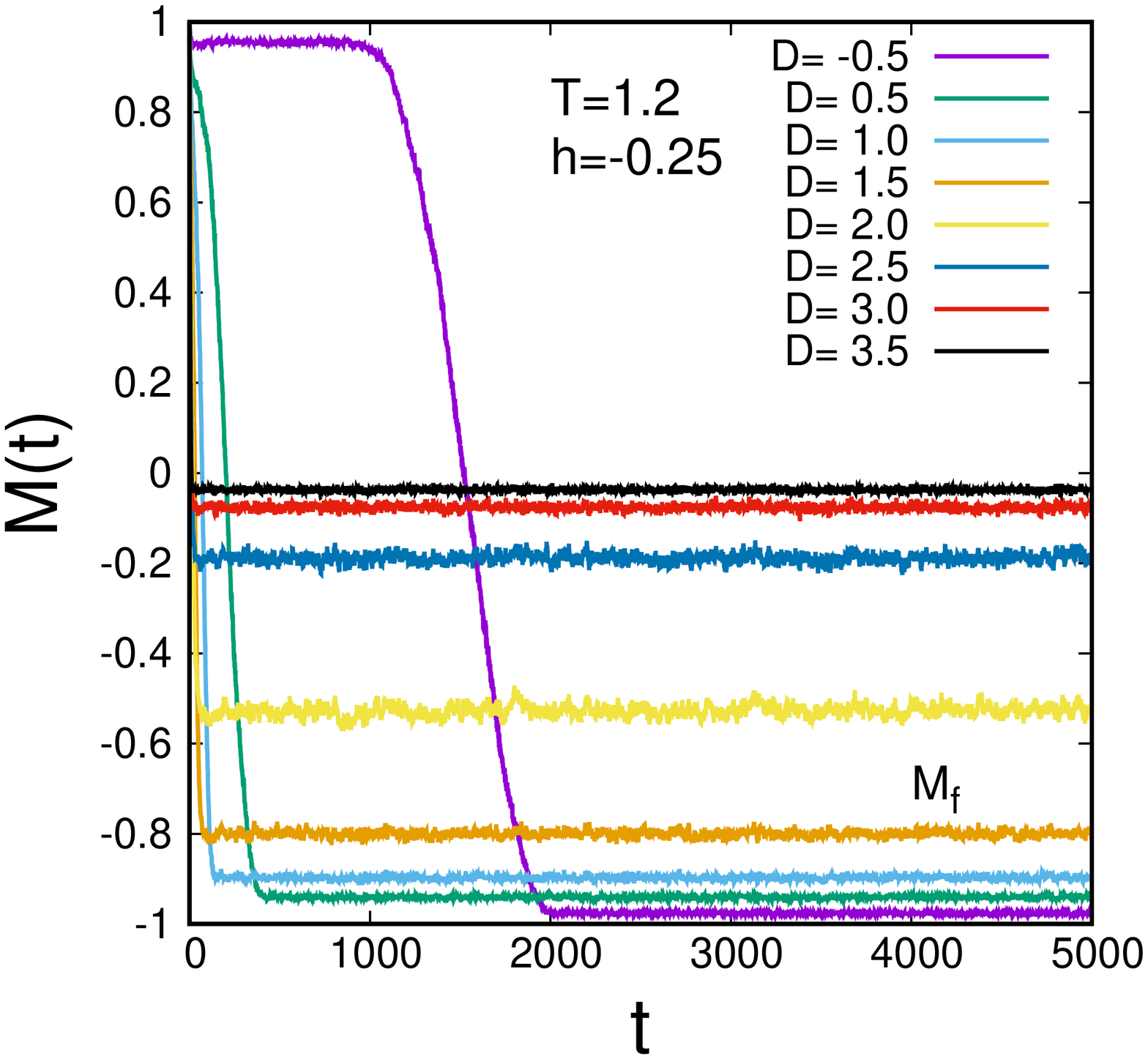}
\caption{Variation  of magnetisation $(M(t))$ with time $(t)$ at different values of 
anisotropy $(D)$ at fixed temperature $T=1.2$ in presence of applied field $h= -0.25$.}
\label{magtime}
\end{center}
\end{figure}

 We have considered that the values of all the spins  are 
$S_i^z=1$, as the initial configuration. This may be imagined that all the spins are in the positive
z-directions of a square lattice in xy-plane. Now a magnetic field 
($h$) in opposite direction (along negative direction of z-axis) is 
applied to the system. That means, $h$ assumes negative value. We have studied the time evolution of the magnetisation ($M(t)$) of the 
system for different strength of anisotropy $(D)$ at any fixed temperature ($T$). 
The '\textit{reversal time}' ($\tau$) is defined as the time by which the 
 magnetisation changes its sign (from initially chosen positive value to the 
 negative value). It is 
observed that reversal time ($\tau$) of the magnetisation decreases with the increase in the strength of anisotropy ($D >0$) (fig-\ref{magtime}). Additionally we noticed that the saturation 
magnetisation ($M_f$), after complete reversal, also varies with the strength
($D$) of the 
anisotropy. $M_f$ is determined by taking time average of the magnetisation after 
reaching saturation (flatness of the plots in negative magnetisation region
in Fig-\ref{magtime}). In the case
 of negative anisotropy $M_f$ reaches a negative value (close to -1) 
i.e. a considerably large number of the spins 
are flipped to $S_i^z=-1$ state (along the direction parallel to the applied 
magnetic field). In the case of positive anisotropy ($D > 0$),
as the magnitude of the  
anisotropy becomes stronger, $|M_f |$ decreases and finally reaches zero. Actually 
for negative $D$ the system behaves as a spin-1/2 model where the system tries to be 
settled down in either of two states (either the spins are 1 or the spins 
are -1) by minimizing its energy. As $D$ is increased beyond $D= -0.5$, the reversal 
time will be large enough. But for positive anisotropy, spins tend to assume 
another value ($S_i^z=0$) favourable for minimizing the energy. 
As a result, the mean density of $S_i^z=0$ starts to 
grow as the positive anisotropy is increased. It is clear that, due to large 
positive ($D>0$) anisotropy, the value of the magnetisation of the system is mostly
determined by $D$ (unlike the situation of negative $D$ where it was preferably
determined by the applied magnetic field $h$).
	
\begin{figure}[h!]
	\begin{subfigure}{0.5\textwidth}
	\includegraphics[angle=-90,width=\textwidth]{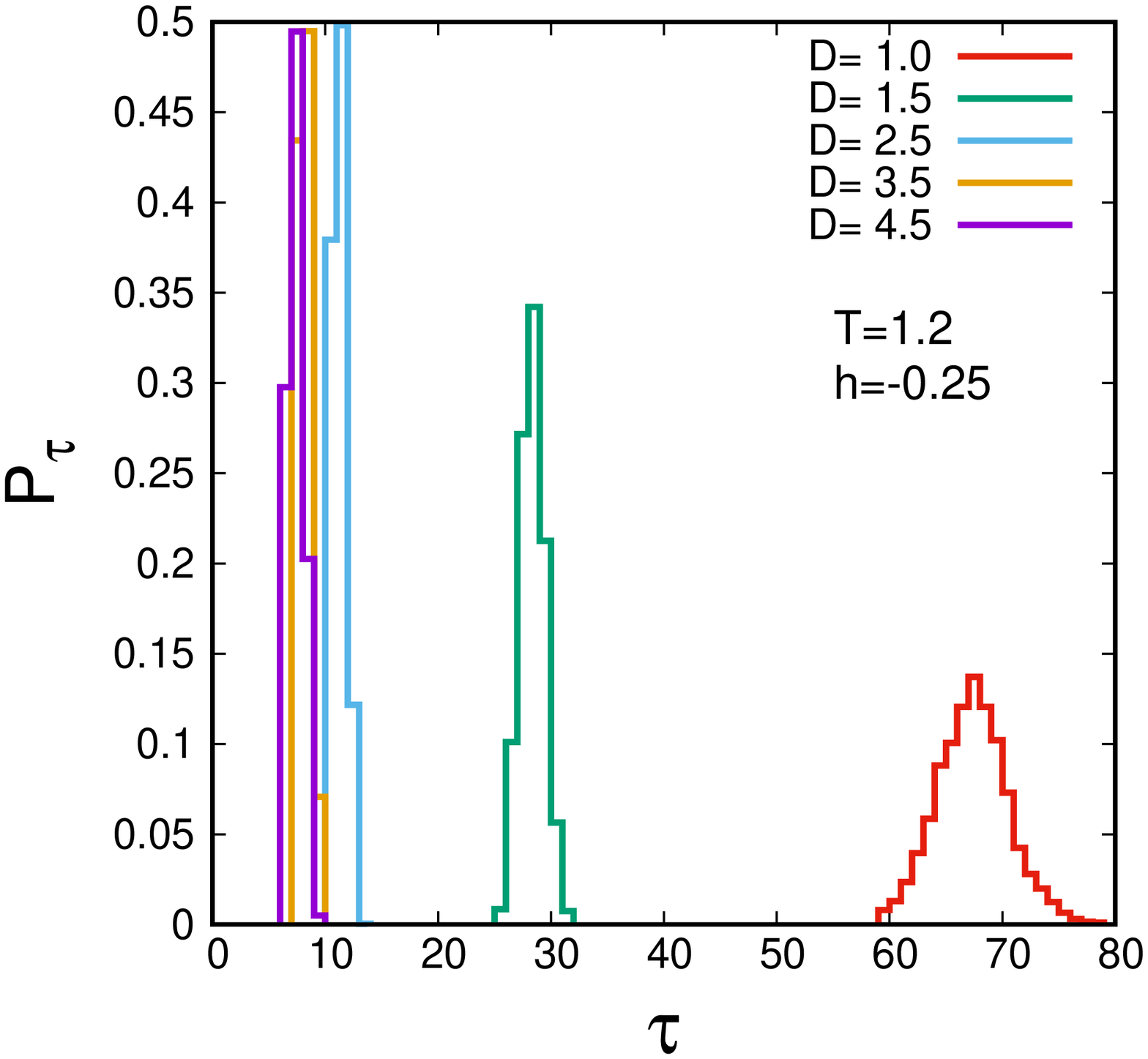}
	\subcaption{}
	\end{subfigure}
	\begin{subfigure}{0.5\textwidth}
	\includegraphics[angle=-90,width=\textwidth]{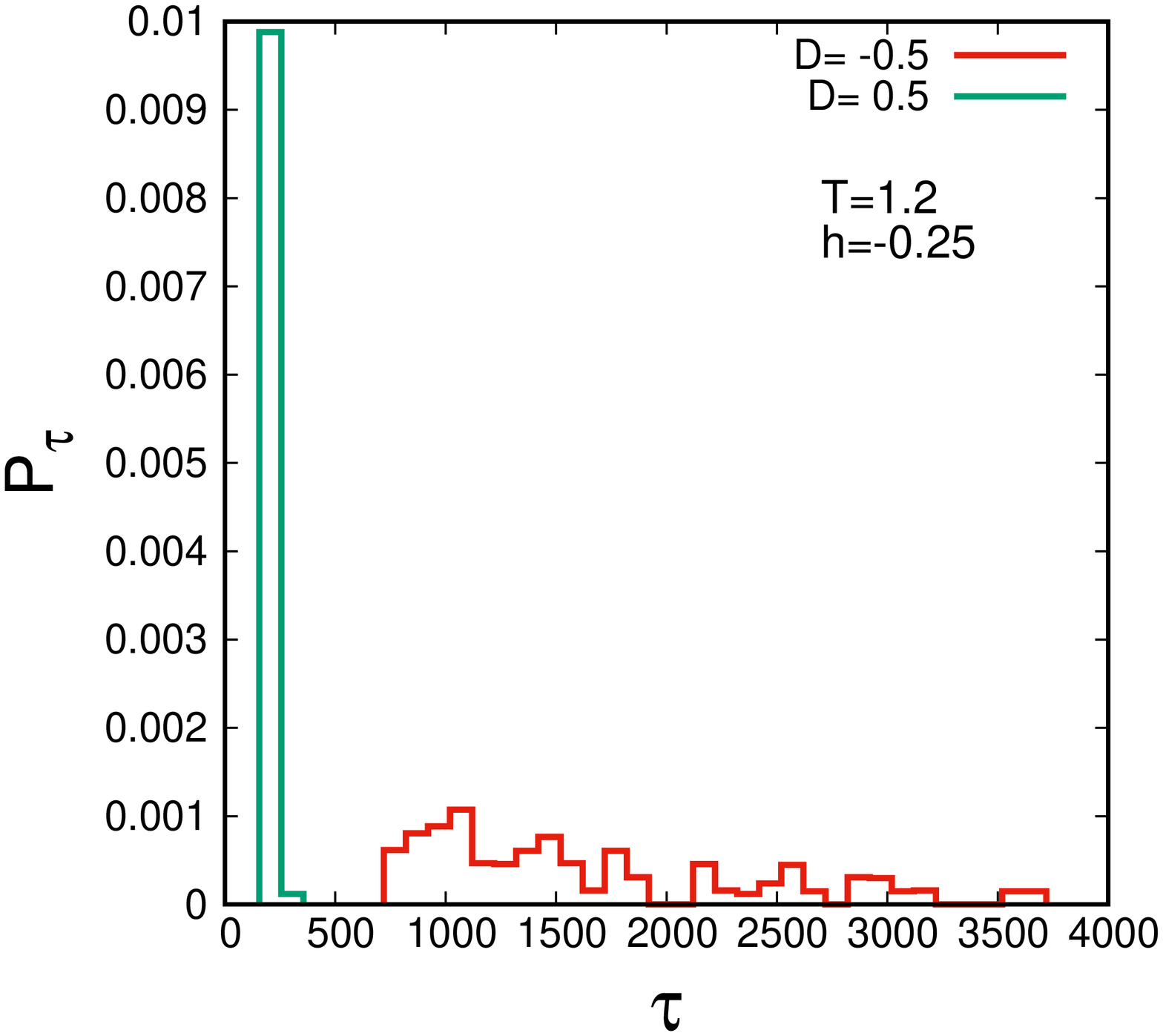}
	\subcaption{}
	\end{subfigure}
\caption{(a) Normalised probability (P$_\tau$) distribution of the reversal 
times ($\tau$) for five different values of positive ($D>0$) anisotrpy at fixed 
temperature $T=1.2$ and applied field $h= -0.25$. (b)  Normalised probability 
distribution of the reversal times for the same absolute value of positive and 
negative  anisotrpy at same temperature and field as (a).}
\label{magtime_dist}
\end{figure}

Fig-\ref{magtime} depicted the reduction in reversal time with the increase in 
anisotropy ($D$) of a single system. Now we have dedicated 10000 number of 
different random samples 
and calculated the reversal time for each. Fig-\ref{magtime_dist} shows normalised 
probability distribution of those reversal times for different strength ($D$) of 
anisotropy. The most probable reversal time and the 
spreading (standard deviation)
of the distribution, decreases with increasing anisotropy
(positive $D$) (fig-\ref{magtime_dist}a). 
Similar study has been carried out and depicted in  fig-\ref{magtime_dist}b 
for both negative and positive anisotropy having same absolute value $|D|=0.5$. 
Distribution for $D= -0.5$ has a huge spread compared to that for $D= +0.5$.  Due to the presence of statistical
distribution of the reversal time ($\tau$), we 
have considered the mean reversal time ($\tau_{av}$) to investigate the
metastable behaviours of this system. 
	
\begin{figure}[h!]
	\begin{subfigure}{0.5\textwidth}
	\includegraphics[angle=-90,width=\textwidth]{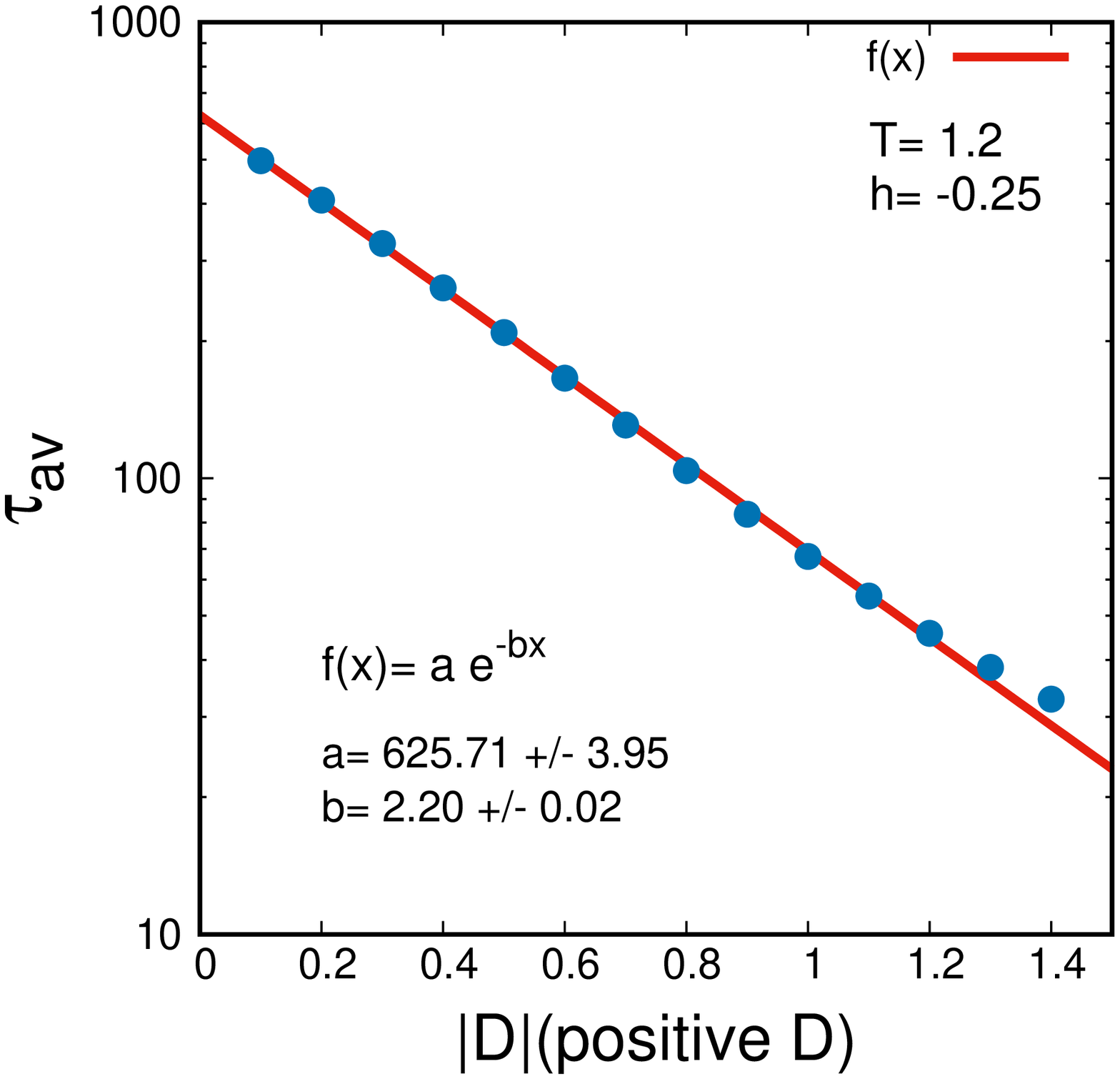}
	\subcaption{}
	\end{subfigure}
	\begin{subfigure}{0.5\textwidth}
	\includegraphics[angle=-90,width=\textwidth]{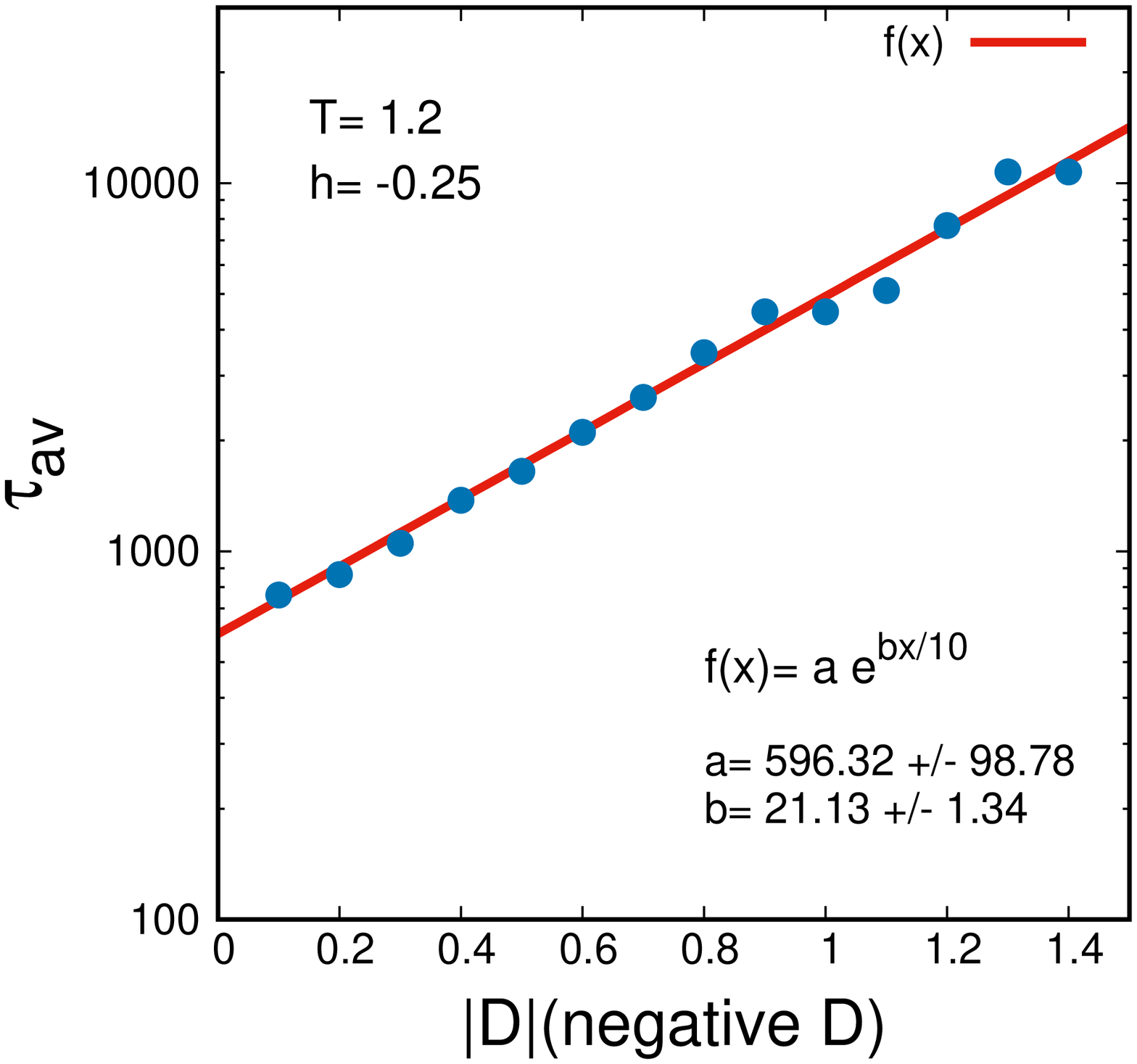}
	\subcaption{}
	\end{subfigure}
\caption{Semilogarithmic plot of variation of mean reversal time ($\tau_{av}$) 
with positive 
anisotropy (a) and negative anisotropy (b) with error or standard 
deviation $\sigma_\tau$ obtained from the 10000 random samples. Data are fitted 
to the function $f(x)= a e^{-bx}$ for positive $D$ and $f(x)= a e^{bx}$ for 
negative $D$ where $f(x)= \tau_{av}$ and $x= |D|$. Temperature is kept $T=1.2$ 
and applied field is $h= -0.25$.}
\label{revtime_d}
\end{figure}

\begin{figure}[h!]
	\begin{subfigure}{0.5\textwidth}
	\includegraphics[angle=-90,width=\textwidth]{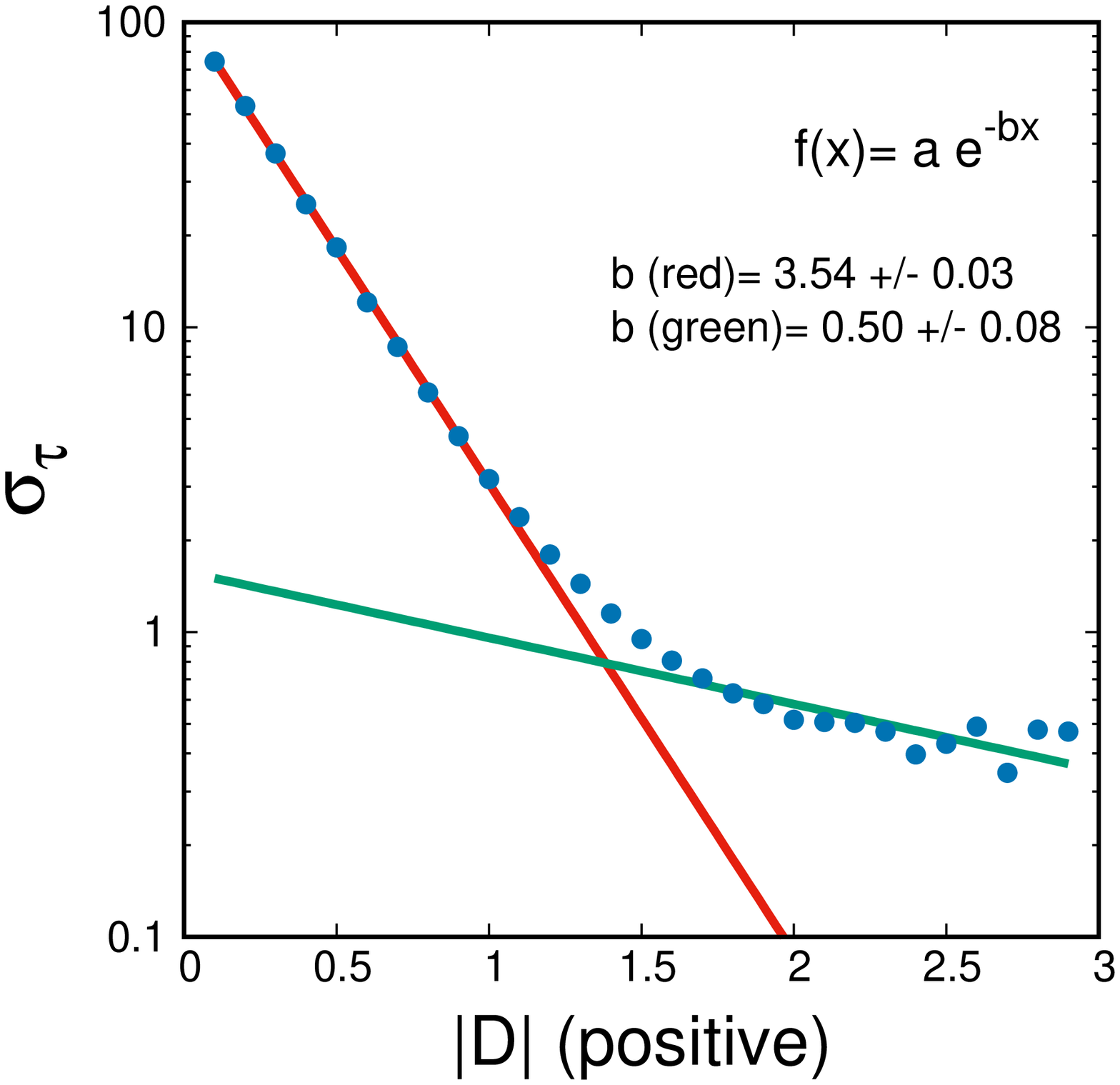}
	\subcaption{}
	\end{subfigure}
	\begin{subfigure}{0.5\textwidth}
	\includegraphics[angle=-90,width=\textwidth]{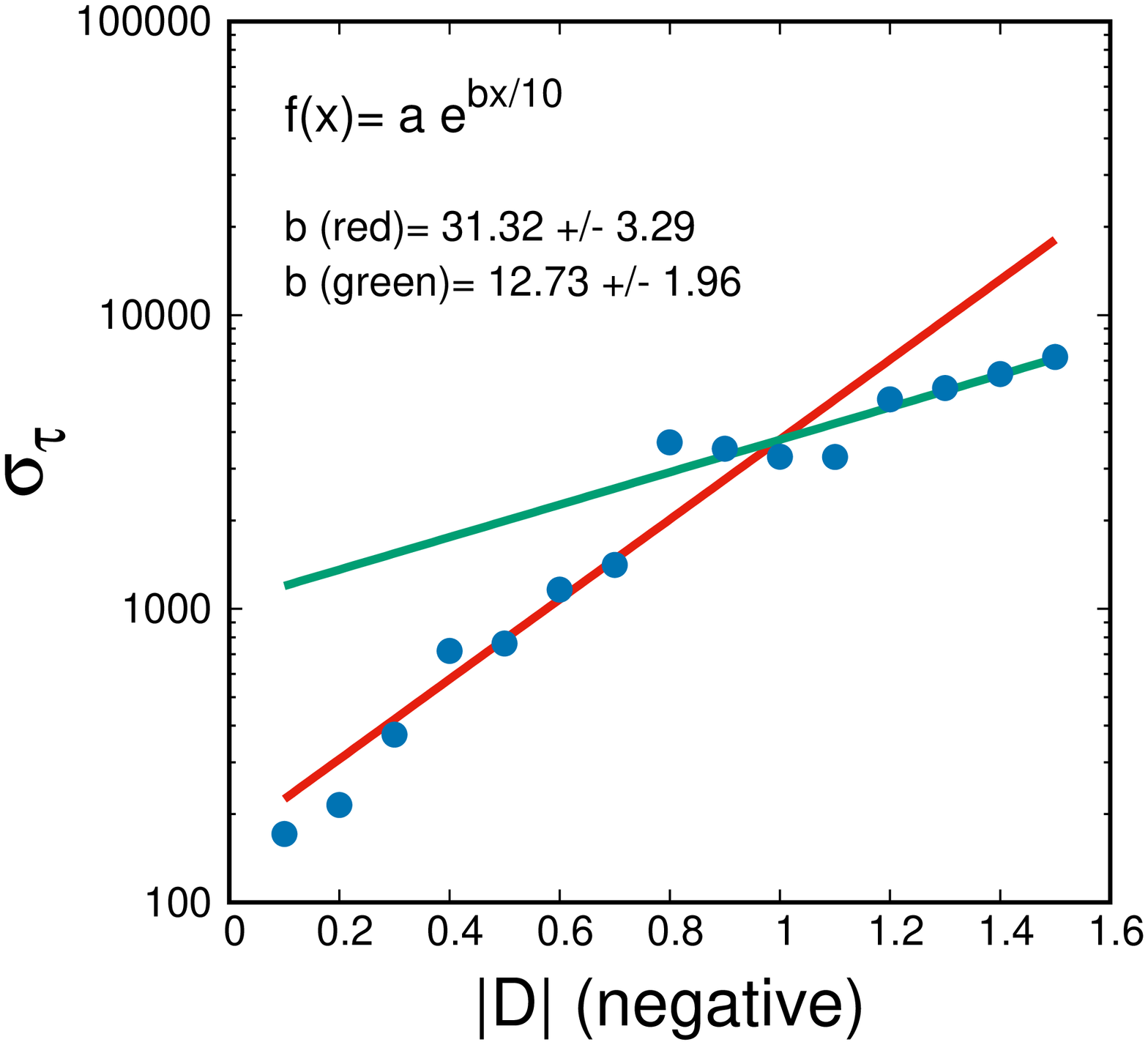}
	\subcaption{}
	\end{subfigure}

\caption{Semilogarithmic plot of the variation of standard deviation 
($\sigma_\tau$) of the mean 
reversal time with positive anisotropy (a) and negative anisotropy. Data 
are fitted to the function $f(x)= a e^{-bx}$ for positive $D$ and 
$f(x)= a e^{bx}$ for negative $D$ where $f(x)= \sigma_\tau$ and $x= |D|$. 
(b)Temperature is kept $T=1.2$ and applied field is $h= -0.25$.}
\label{revtimesd_d}
\end{figure}

 We have studied the nature of the variation of the mean reversal time 
($\tau_{av}$) with anisotropy (for both positive $D$ fig-\ref{revtime_d}a and negative 
$D$ fig-\ref{revtime_d}b). Mean reversal time is calculated by simply averaging the 
reversal times obtained for 10000 numbers of different random samples.
The mean $\tau_{av}$ of the reversal times and its standard 
deviation $\sigma_{\tau}$ (fig-\ref{revtimesd_d}a, \ref{revtimesd_d}b) were
found to decrease 
exponentially ($\tau_{av} \sim e^{-gD}$) with the increase in positive anisotropy. In the presence of negative 
anisotropy, both of them increase exponentially ($\tau_{av} \sim e^{-g'D}$) with the increase in 
the absolute value of the strength of anisotropy. 
Positive stronger anisotropy compel the system to lower the absolute value of magnetisation, due to the production of large number of $S_i^z=0$ (which contributes
nothing to the magnetisation). Stronger value of negative $D$ will map the system
onto an equivalent spin-${{1} \over {2}}$ Ising ferromagnet, where the single spin
flip would require more cost of energy than that of a Blume-Capel ferromagnet with
positive $D$, which has a possibility of transition from $S_i^z=+1$ to $S_i^z=0$.
This is a possible reason of getting smaller reversal time in the case of larger
positive $D$, in the BC model. In both cases (i.e., $D>0$ and $D<0$), a crossover
is observed in the rate ($r'$) of change of the exponential function of $\sigma_{\tau}
\sim e^{-r'D}$.

\begin{figure}[h!]
\centering
	\begin{subfigure}{0.45\textwidth}
	\includegraphics[angle=-90,width=\textwidth]{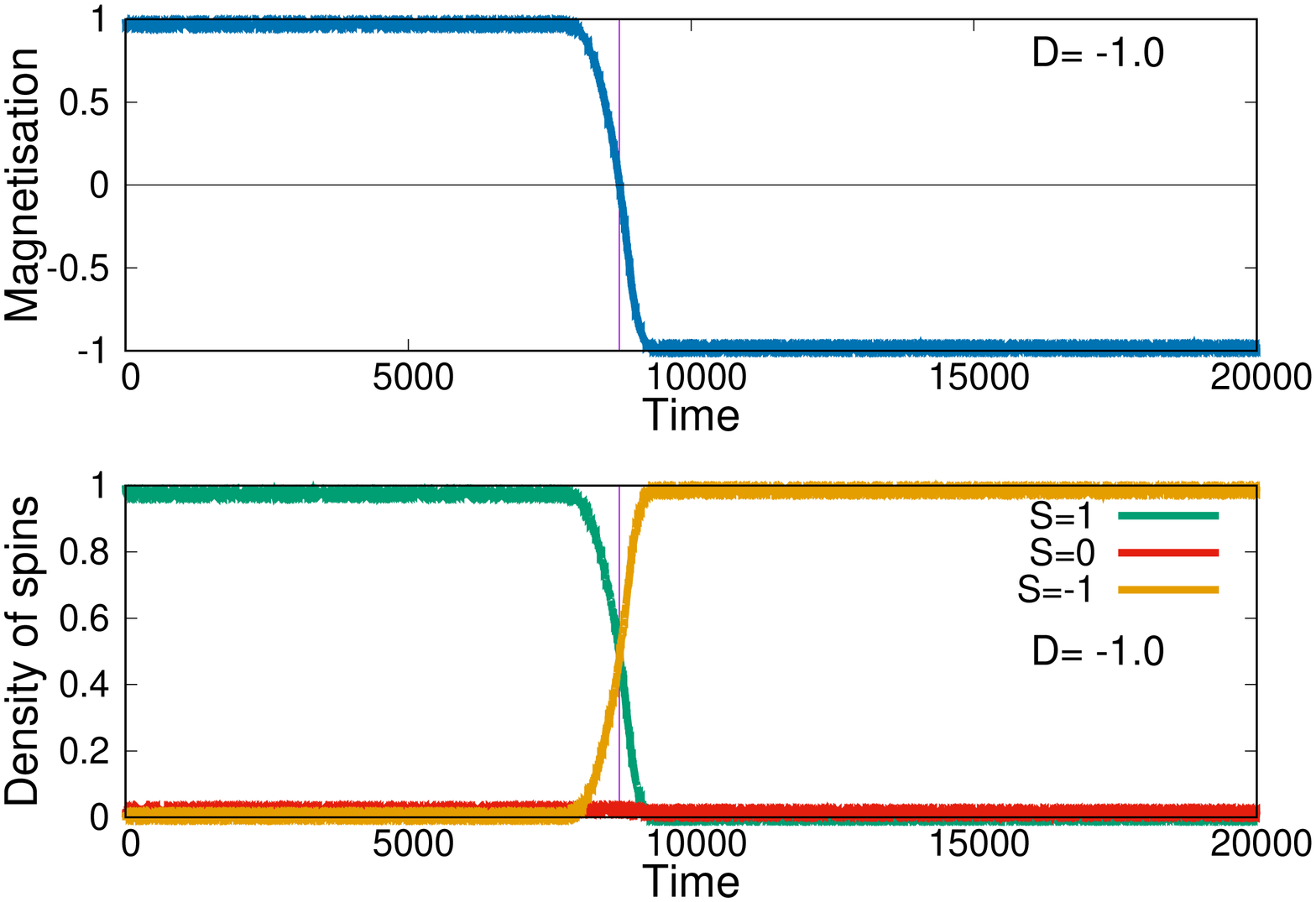}
	\subcaption{}
	\end{subfigure}
	\begin{subfigure}{0.45\textwidth}
	\includegraphics[angle=-90,width=\textwidth]{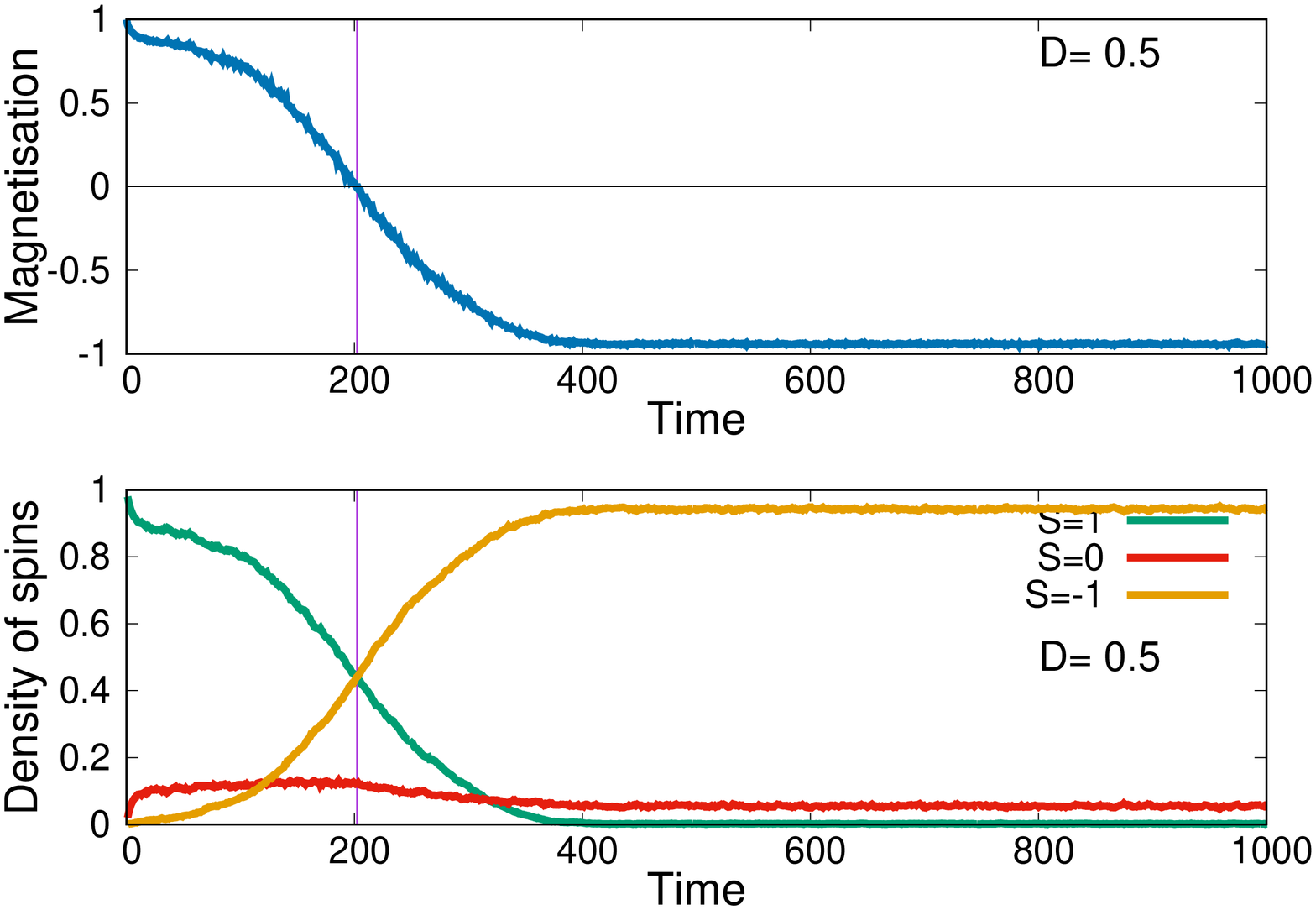}
	\subcaption{}
	\end{subfigure}
	\begin{subfigure}{0.45\textwidth}
	\includegraphics[angle=-90,width=\textwidth]{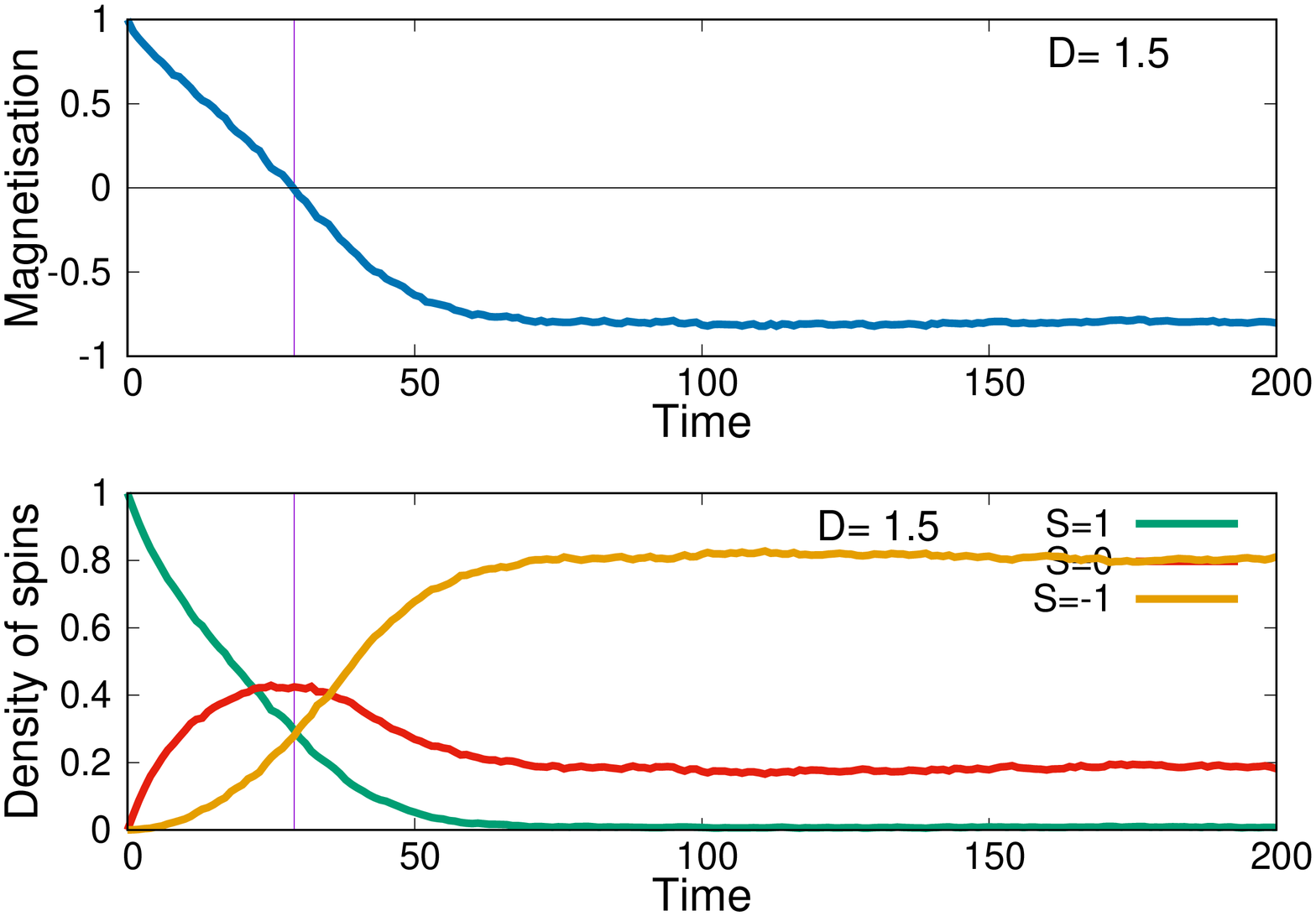}
	\subcaption{}
	\end{subfigure}
	\begin{subfigure}{0.45\textwidth}
	\includegraphics[angle=-90,width=\textwidth]{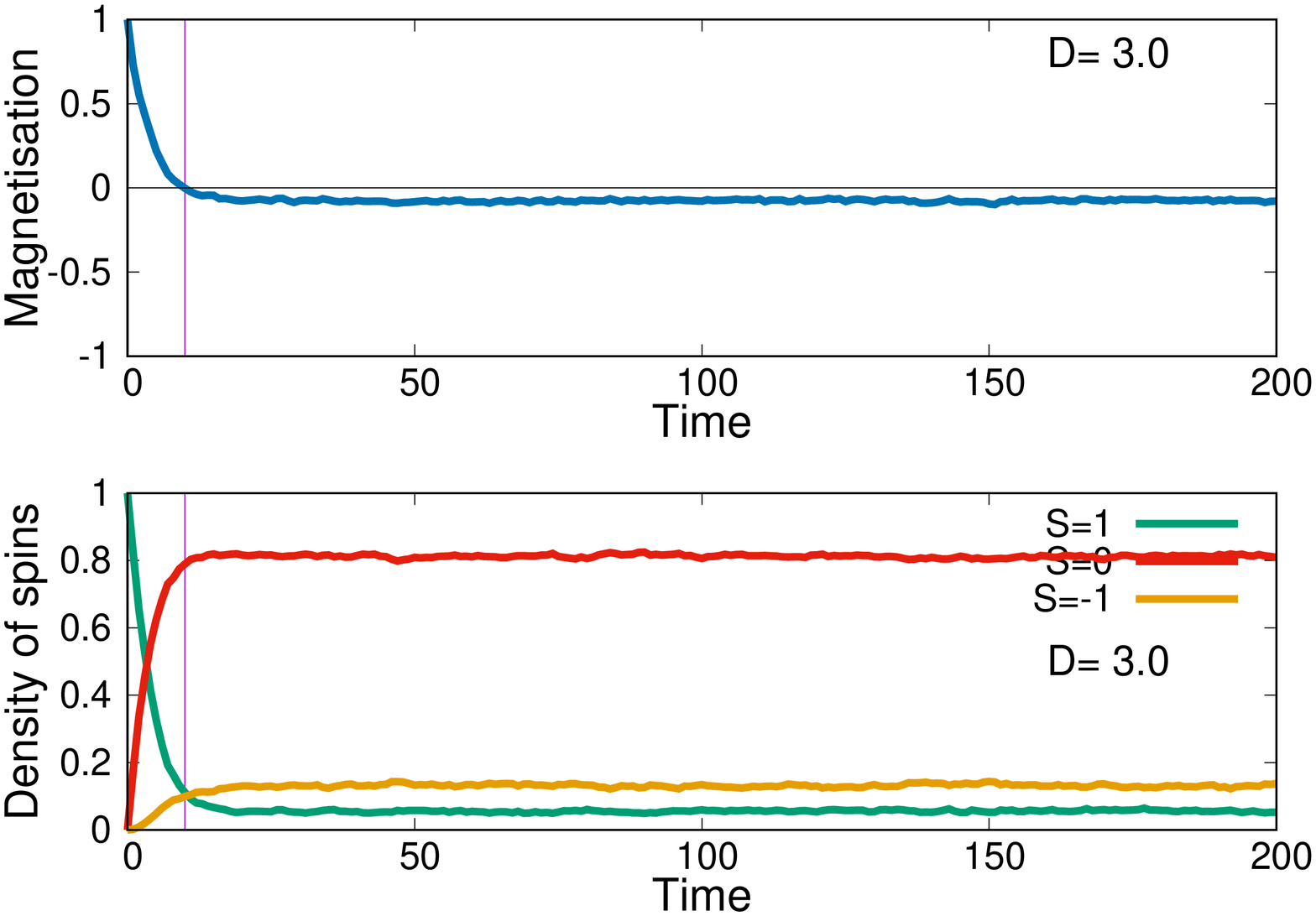}
	\subcaption{}
	\end{subfigure}
\caption{Variations of the mean densities of $S_i^z=+1$, $S_i^z=0$ and
$S_I^z=-1$ with time for different 
values of anisotropy ($D$) at fixed values of the temperature ($T=1.2$) 
and the applied field 
($h= -0.25$). Vertical lines in each figure indicates the time of reversal.}
\label{spindyn}
\end{figure}

\begin{figure}[h!]
	\begin{subfigure}{0.333\textwidth}
	\includegraphics[angle=-90,width=\textwidth]{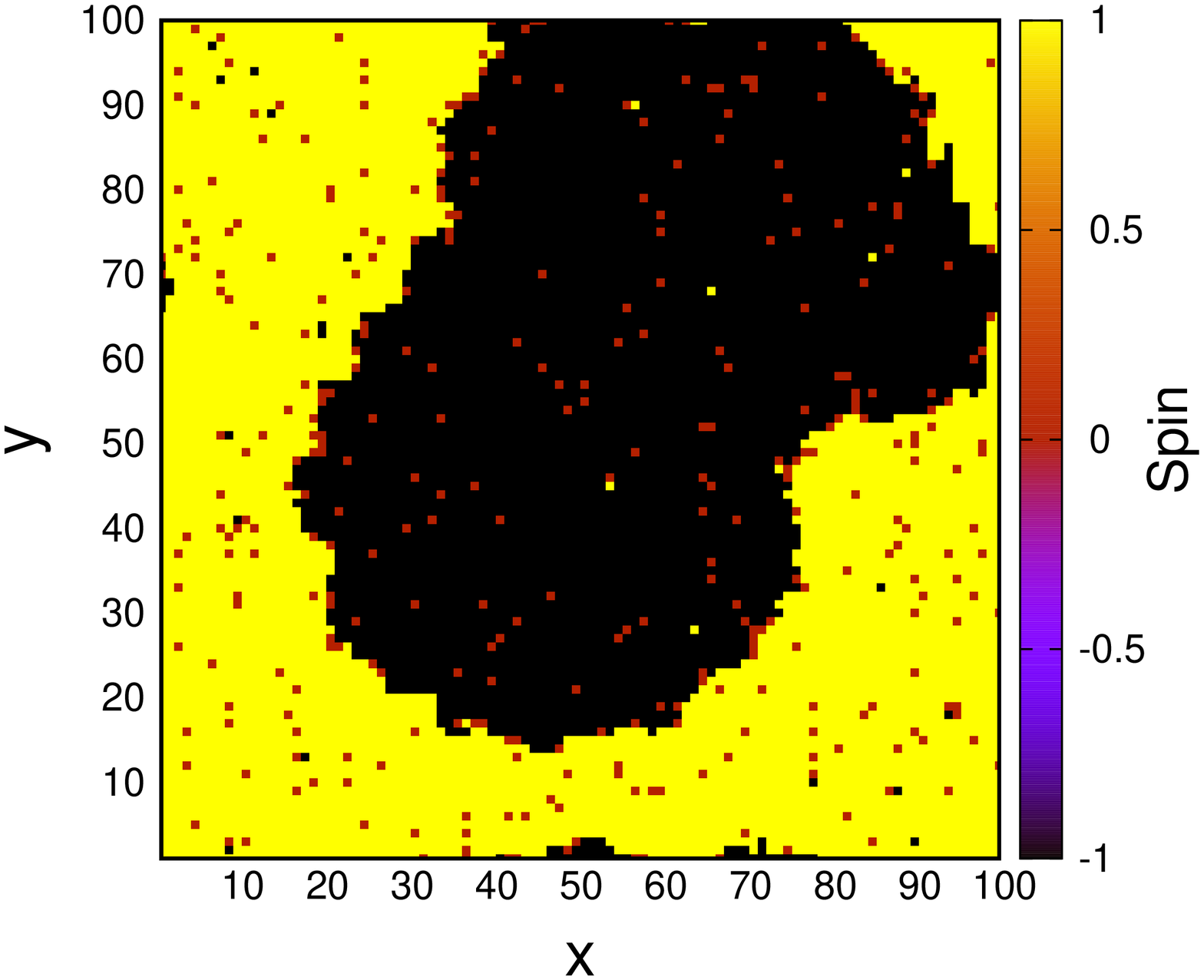}
	\subcaption{}
	\end{subfigure}
	\begin{subfigure}{0.333\textwidth}
	\includegraphics[angle=-90,width=\textwidth]{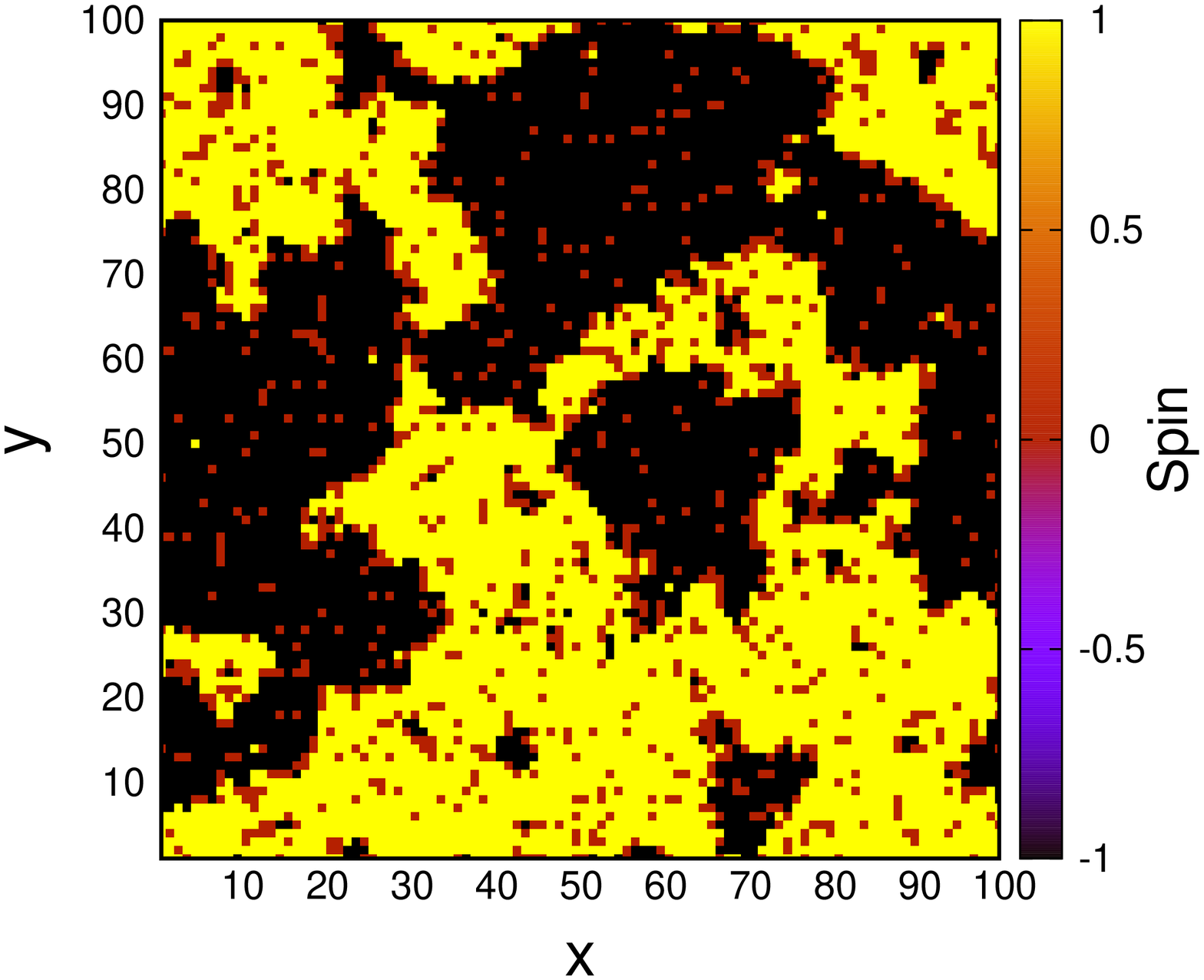}
	\subcaption{}
	\end{subfigure}
	\begin{subfigure}{0.333\textwidth}
	\includegraphics[angle=-90,width=\textwidth]{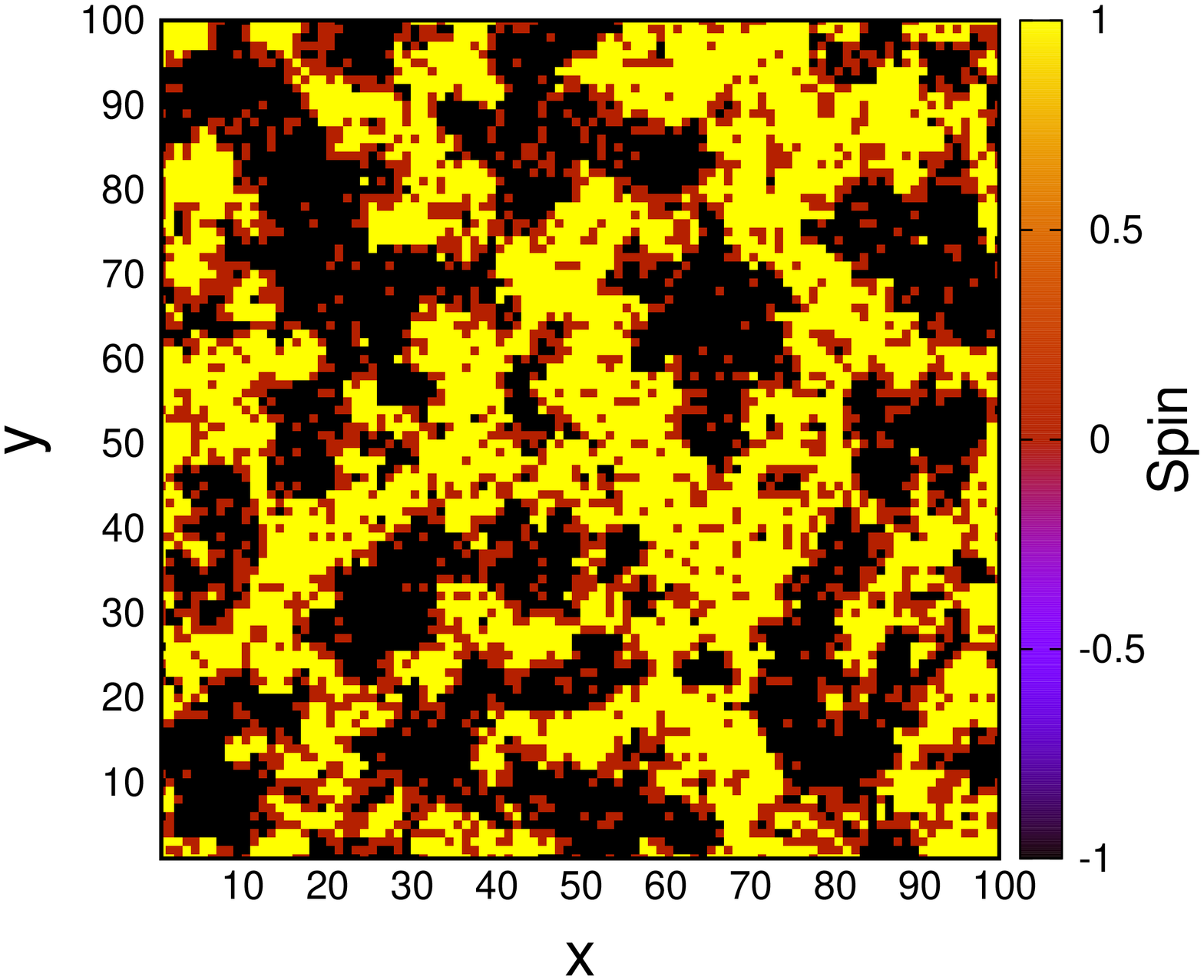}
	\subcaption{}
	\end{subfigure}
	\begin{subfigure}{0.333\textwidth}
	\includegraphics[angle=-90,width=\textwidth]{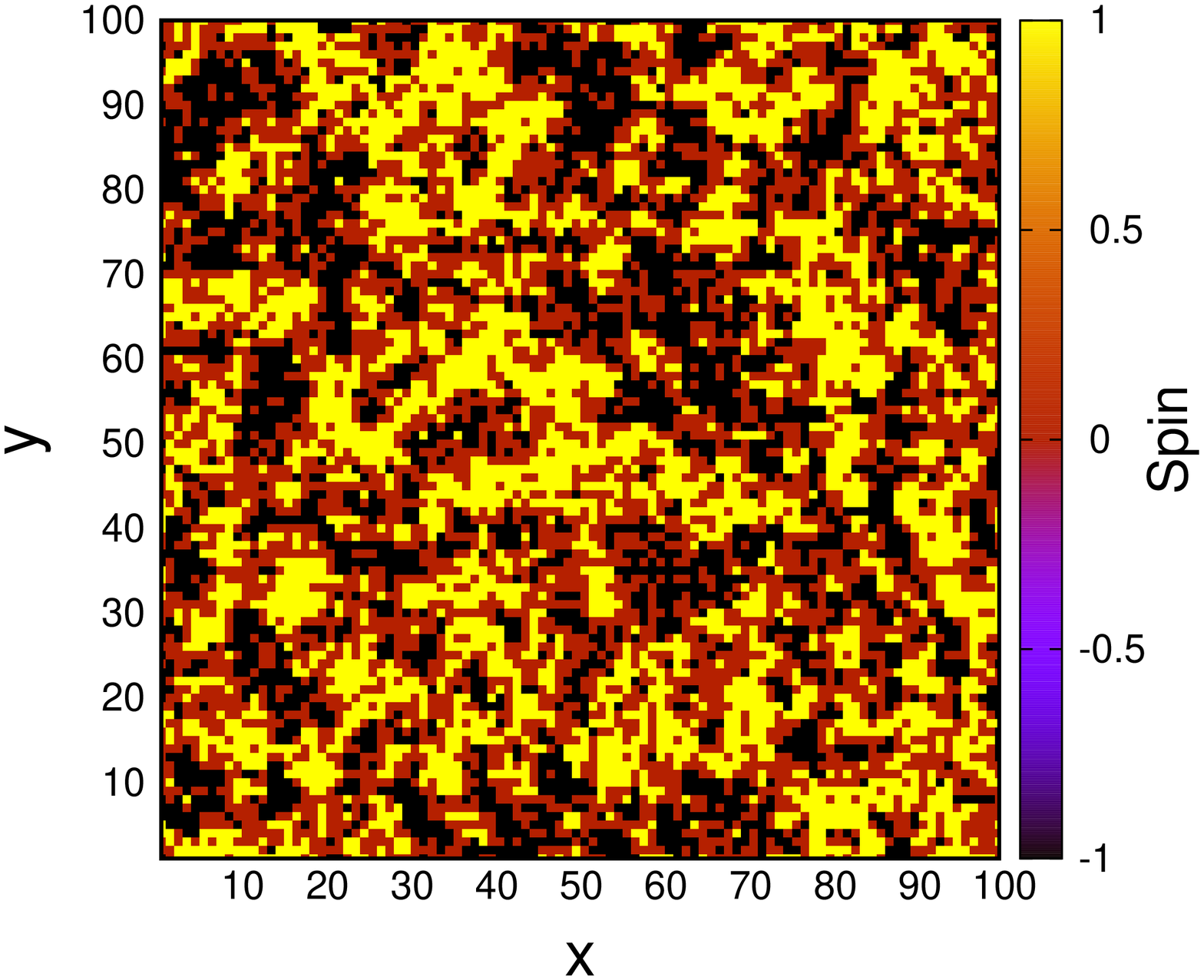}
	\subcaption{}
	\end{subfigure}
	\begin{subfigure}{0.333\textwidth}
	\includegraphics[angle=-90,width=\textwidth]{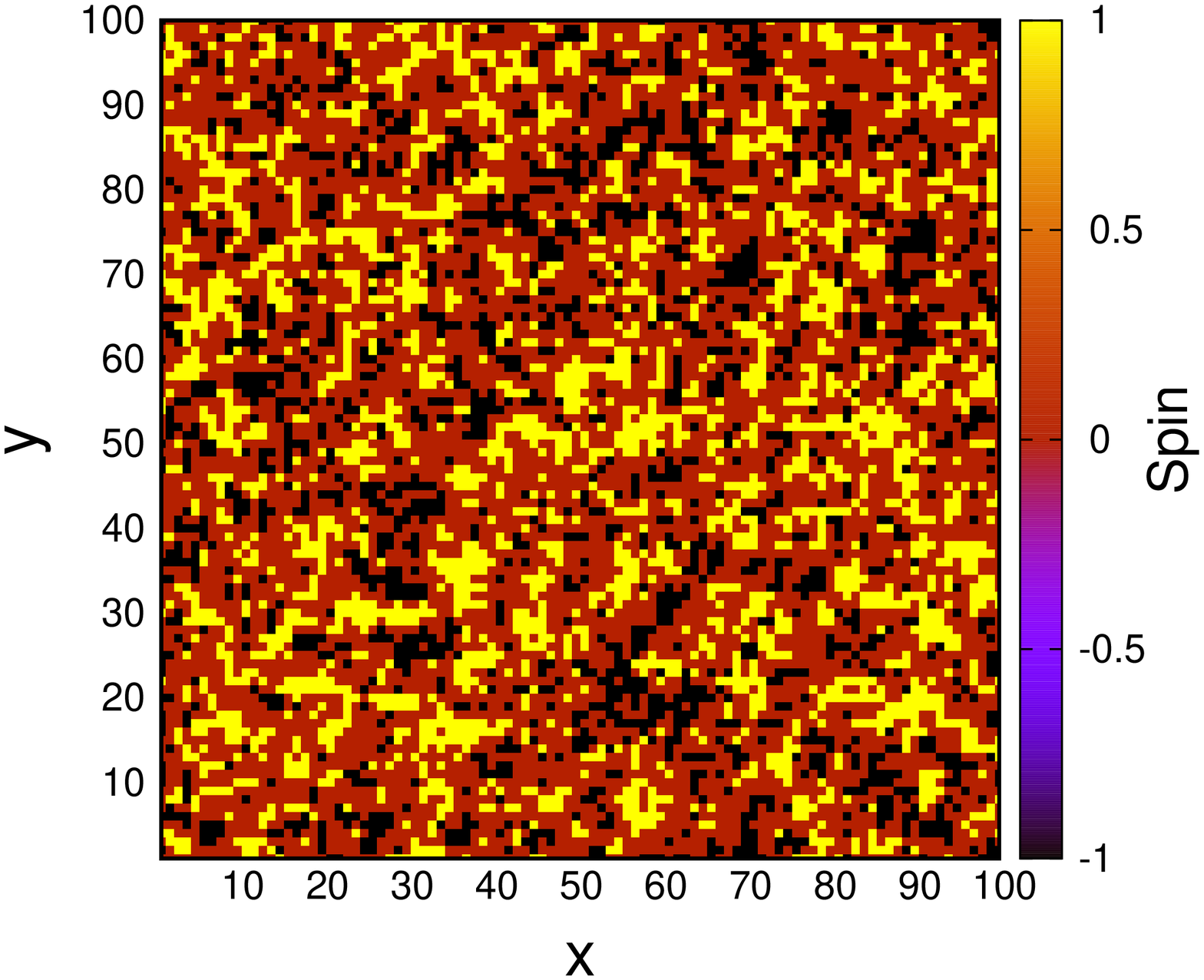}
	\subcaption{}
	\end{subfigure}
	\begin{subfigure}{0.333\textwidth}
	\includegraphics[angle=-90,width=\textwidth]{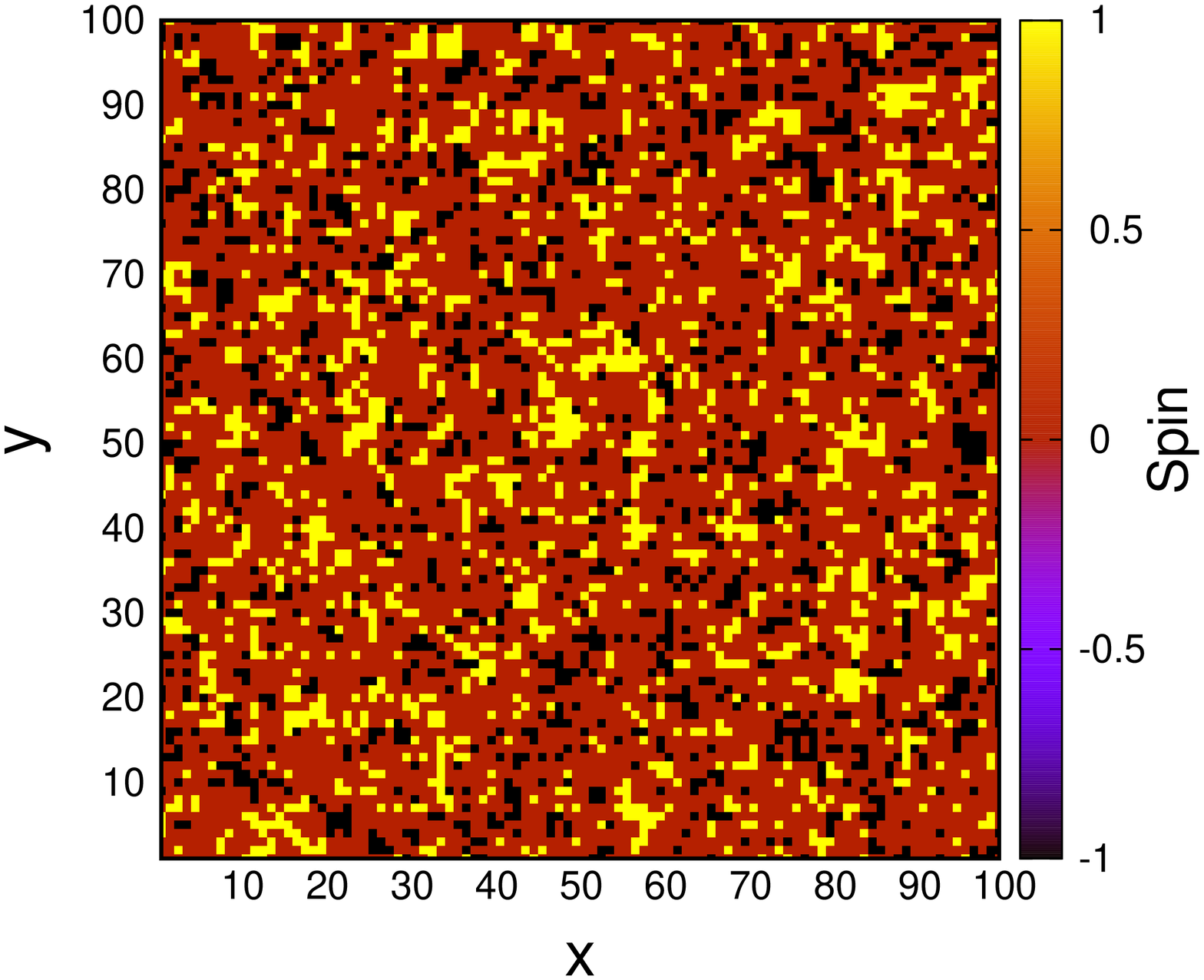}
	\subcaption{}
	\end{subfigure}
\caption{Snapshots of the spin configurations captured at the time 
of reversal for the different values of the strength of 
anisotropy $D$. (a) $D= -0.5$ and 
$\tau= 1516$ MCSS (b)  $D= 0.5$ and $\tau= 201$ MCSS (c)  $D= 1.0$ and 
$\tau= 69$ MCSS (d)  $D= 1.5$ and $\tau= 28$ MCSS (e)  $D= 2.0$ and 
$\tau= 16$ MCSS (f)  $D= 2.5$ and $\tau= 12$ MCSS. 
Applied field is fixed at $h= -0.25$.}
\label{snaps1}
\end{figure}

To understand the reason, of the above mentioned observations, clearly, we have investigated the dynamics
of the mean density 
(fig-\ref{spindyn}) of the three values of $S_i^z$, i.e.,+1,0 and -1 for different values of $D$. For 
negative ($D<0$) anisotropy, the mean density of $S_i^z=+1$ is almost fixed 
and maximum (approximately equal to unity) in the 
metastable region and suddenly starts to decrease near reversal time and 
vanishes after some time (as a result of complete reversal dominated by large
population of $S_i^z=-1$). As a result, the density of $S_i^z=-1$ remains zero in the 
metastable region then suddenly acquires a large value (due to reversal) and 
eventually acquires maximum value when complete reversal is achieved. 
The mean density of $S_i^z=0$
remains almost zero through out all the time in 
presence of negative anisotropy ($D<0$). This fact arises due to the 
behaviour of the BC model which
is similar to that of Ising model in the case of large negative anisotropy. 
As the positive anisotropy rises, the active role of 
$S_i^z=0$ comes into the picture. This shows a maximum near the reversal 
time for some 
moderate values 
of anisotropy ($D>0$). 
Some snapshots (fig-\ref{snaps1}) of instantaneous spin configurations, are
 captured near the time of 
reversal in the presence of different anisotropy ($D>0$) to have a clear idea. From 
fig-\ref{spindyn} it is observed 
that, close to the time of reversal of the magnetisation, the 
density of $S_i^z=+1$ and $S_i^z=-1$ are equal and the density of $S_i^z=0$ 
gets peaked 
(for moderate value of positive $D$) or a larger value (for higher $D$). This is obviously true because 
for vanishingly small net magnetisation, the densities of $S_i^z=+1$ and $S_i^z=-1$
must be approximately equal, in the thermodynamic limit. The values of 
this equal density (of $S_i^z=+1$ and $S_i^z=-1$) and the density of the 
$S_i^z=0$ are legitimately noticed to be varied with anisotropy ($D>0$).  
	
\begin{figure}[htb]
	\begin{subfigure}{0.5\textwidth}
	\includegraphics[angle=-90,width=\textwidth]{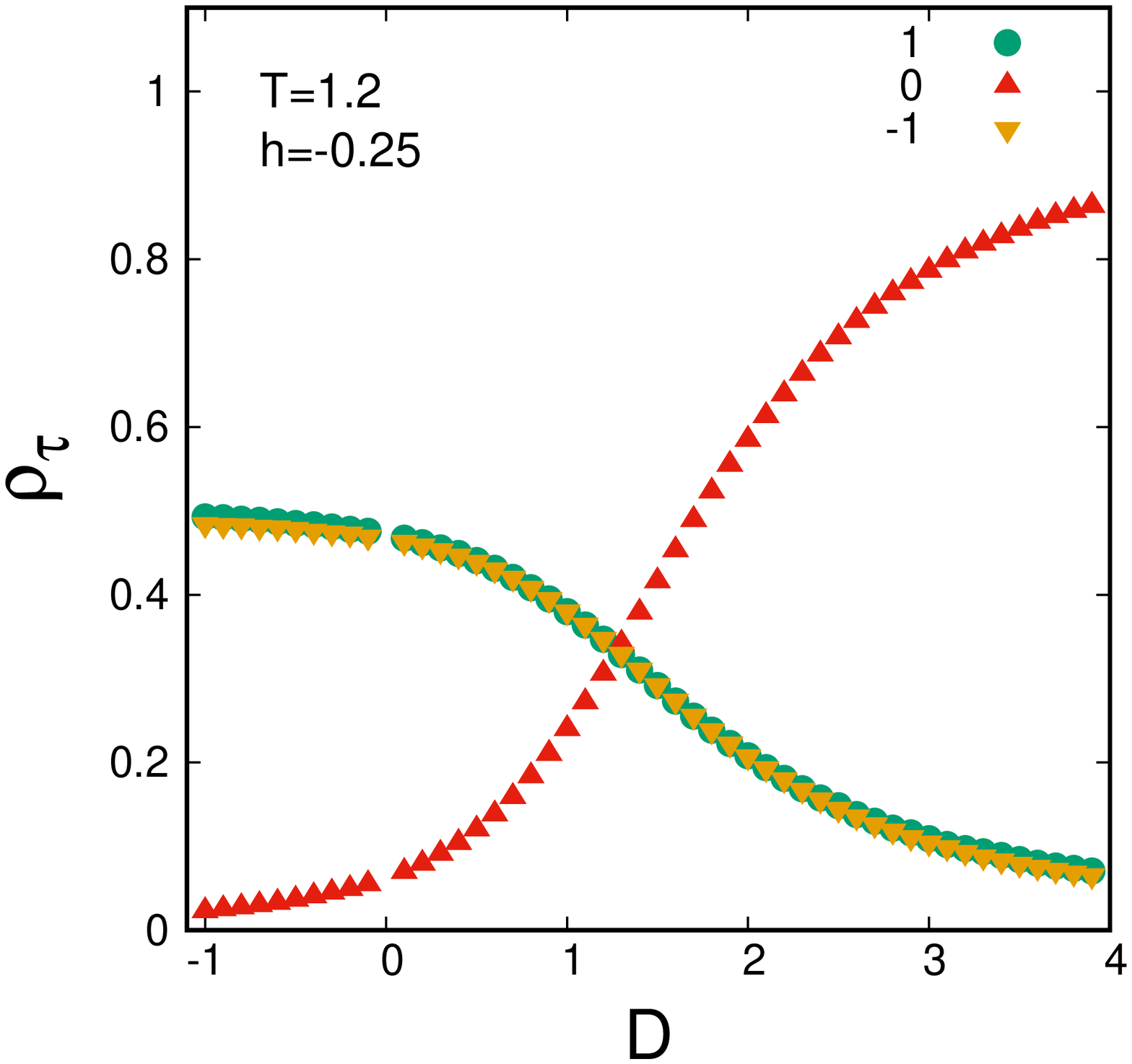}
	\subcaption{}
	\end{subfigure}
	\begin{subfigure}{0.5\textwidth}
	\includegraphics[angle=-90,width=\textwidth]{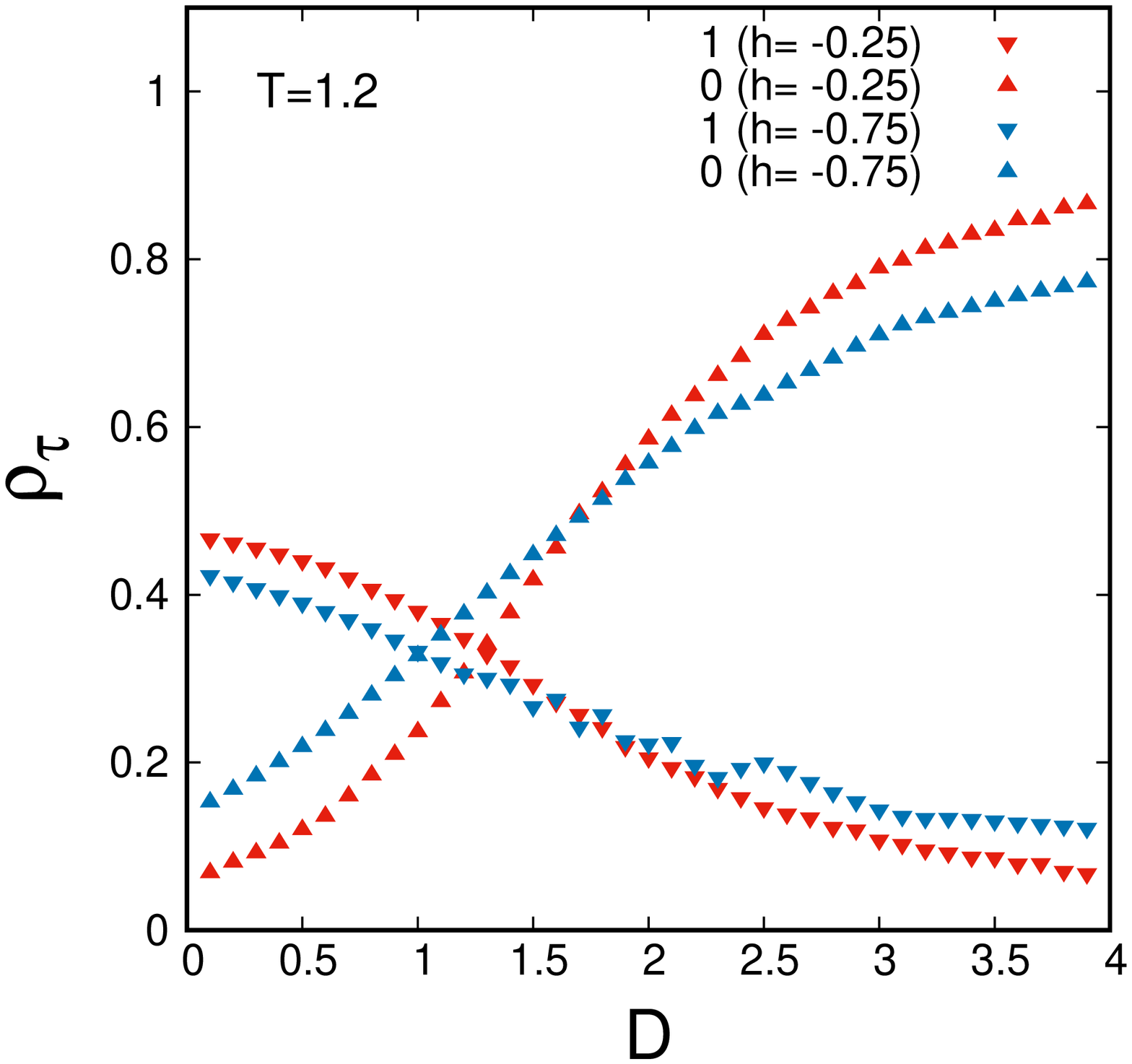}
	\subcaption{}	
	\end{subfigure}
\caption{(a) Variations of the mean densities of $S_i^z=+1$, $S_i^z=0$ and $S_i^z=-1$
at the 
time of reversal ($\rho_\tau$) with fixed anisotropy $(D)$. Temperature is set 
to $T=1.2$ and the applied field is $h= -0.25$ (b) Variation of $\rho_\tau$ 
with $D$ in the presence of two different values of fields $h= -0.25$ (red) 
and $h= -0.75$ (blue).Temperature is fixed to $T=1.2$.}
\label{revtime_density}
\end{figure}

If we plot (fig-\ref{revtime_density}a) the densities($\rho_\tau$) of these three values
(+1, 0 and -1) of
$S_i^z$ in the vicinity of the 
time of reversal ($\tau$) with anisotropy ($D>0$), it is observed that the densities 
of $S_i^z=+1$ and $S_i^z=-1$ reduce with increasing anisotropy whereas the density of $S_i^z=0$ increases. 
Interestingly, a strength of anisotropy has been noticed, in presence of which 
the three values (+1, 0 and -1) of $S_i^z$ 
 have almost equal density near the time of reversal. Since, the stronger 
 field ($h$) accelerates the reversal process for a particular value of 
anisotropy ($D$), the $\rho_\tau$ of $S_i^z=+1$ and $S_i^z=-1$ are reduced 
and that for $S_i^z= 0$ is increased. So 
obviously that particular value of $D(>0)$, 
for which $\rho_\tau$ of all $S_i^z=+1$, $S_i^z=0$ and $S_i^z=-1$ are equal, is reduced 
in presence of a stronger applied field (fig-\ref{revtime_density}b). For clarity,
 in 
fig-\ref{revtime_density}b we have excluded the density of $S_i^z=-1$ as it is same as 
that of $S_i^z=+1$.  

\begin{figure}[h!]
\centering
	\includegraphics[angle=-90,width=0.5\textwidth]{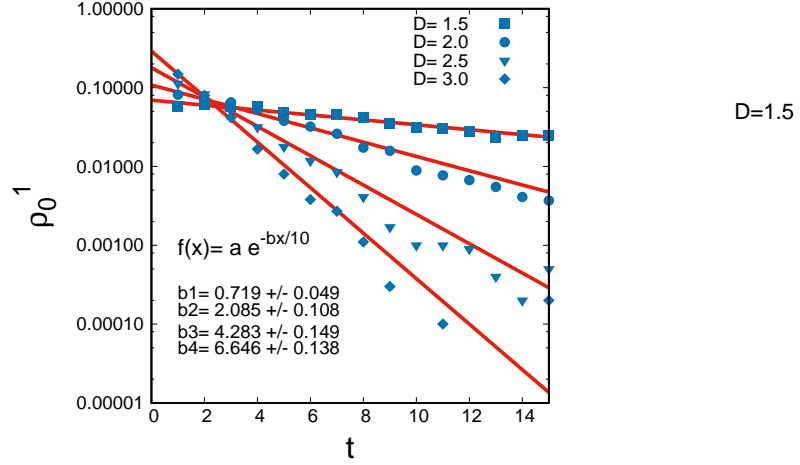}
\caption{Temporal evolution of the density of $S_i^z=0$ surrounded by all 
neighbouring four $S_i^z=+1$  ($\rho_0^1$) for four different set of values 
of anisotropy $D$.
(Here, $D= 1.5$, $D= 2.0$, $D= 2.5$ and $D= 3.0$ ). 
All the four plots are fitted to the function $f(x)= a e^{-bx/10}$ 
where $f(x)$ stands for $\rho_0^1$ and $x$ stands for $t$. Applied field is $h= -0.25$ and $T=1.2$.}
\label{revtime_confirm1}
\end{figure}

\begin{figure}[h!]
	\begin{subfigure}{0.333\textwidth}
	\includegraphics[angle=-90,width=\textwidth]{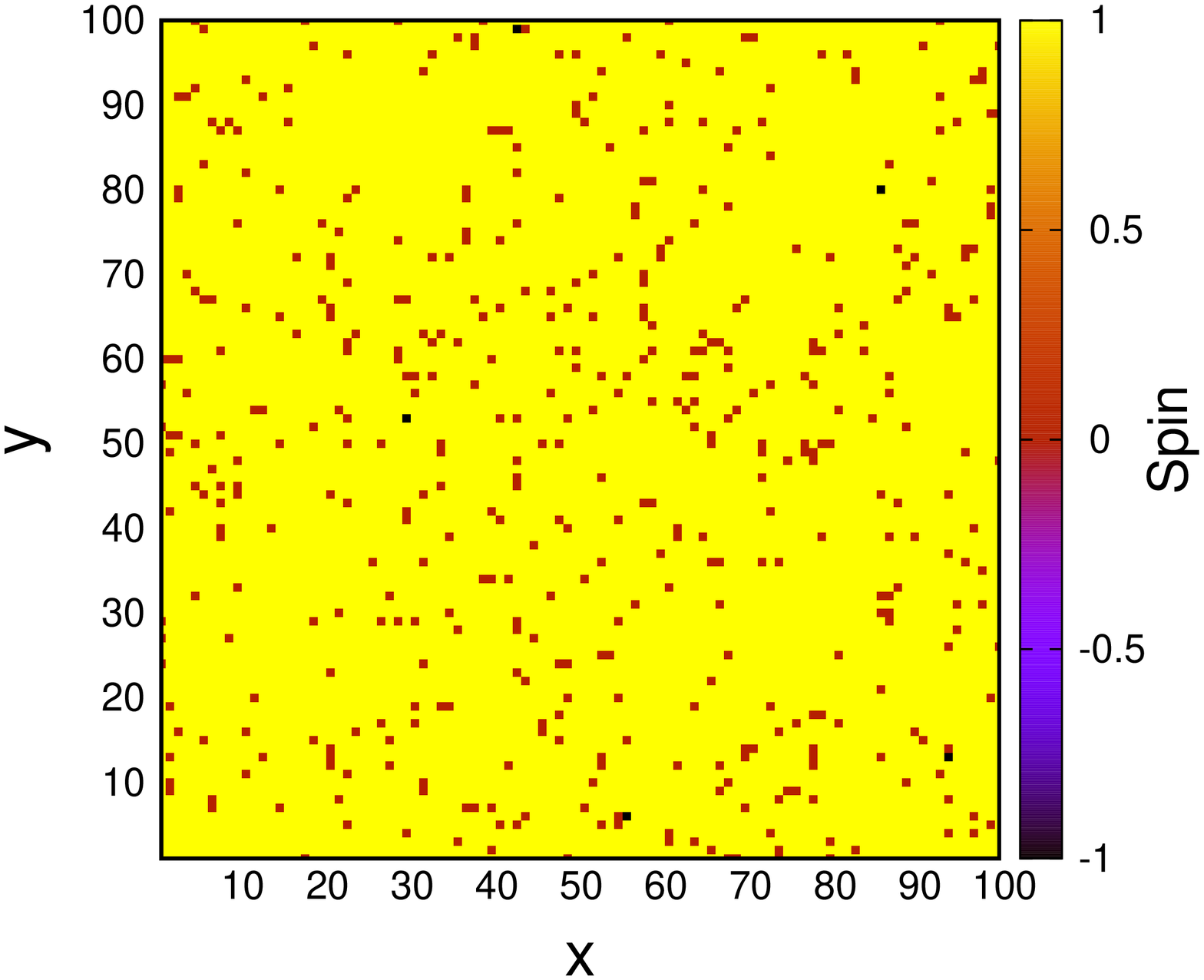}
	\subcaption{}
	\end{subfigure}
	\begin{subfigure}{0.333\textwidth}
	\includegraphics[angle=-90,width=\textwidth]{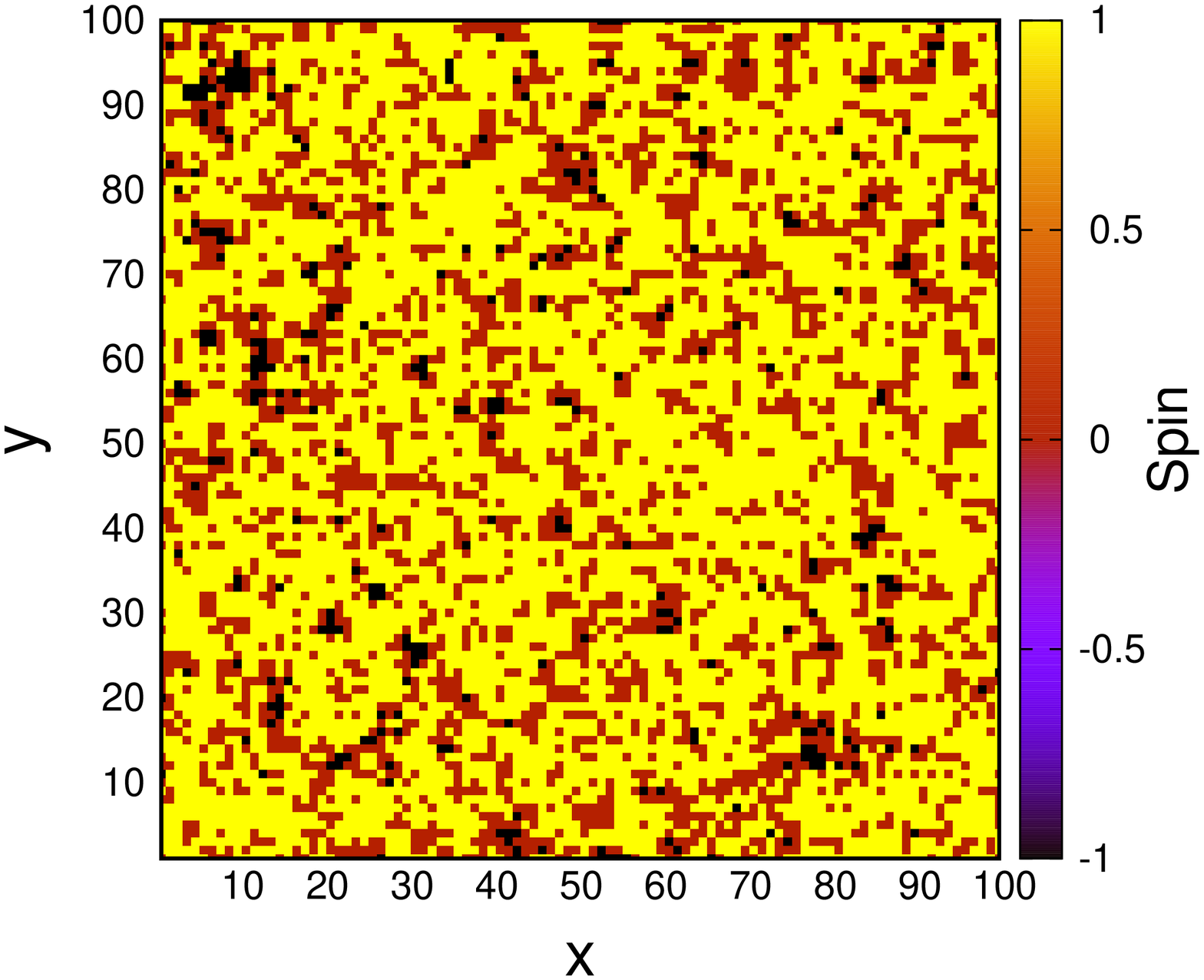}
	\subcaption{}
	\end{subfigure}
	\begin{subfigure}{0.333\textwidth}
	\includegraphics[angle=-90,width=\textwidth]{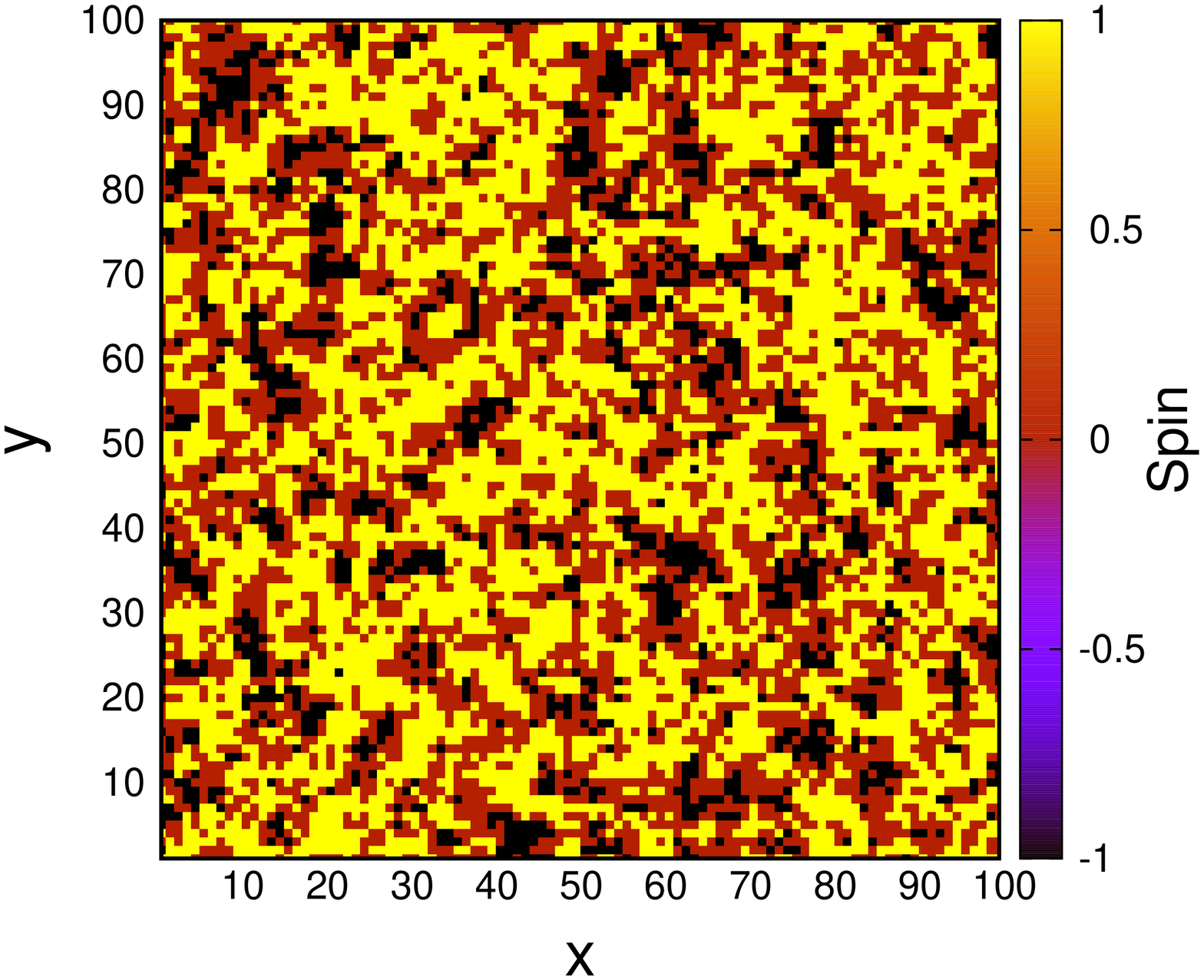}
	\subcaption{}
	\end{subfigure}
	\begin{subfigure}{0.333\textwidth}
	\includegraphics[angle=-90,width=\textwidth]{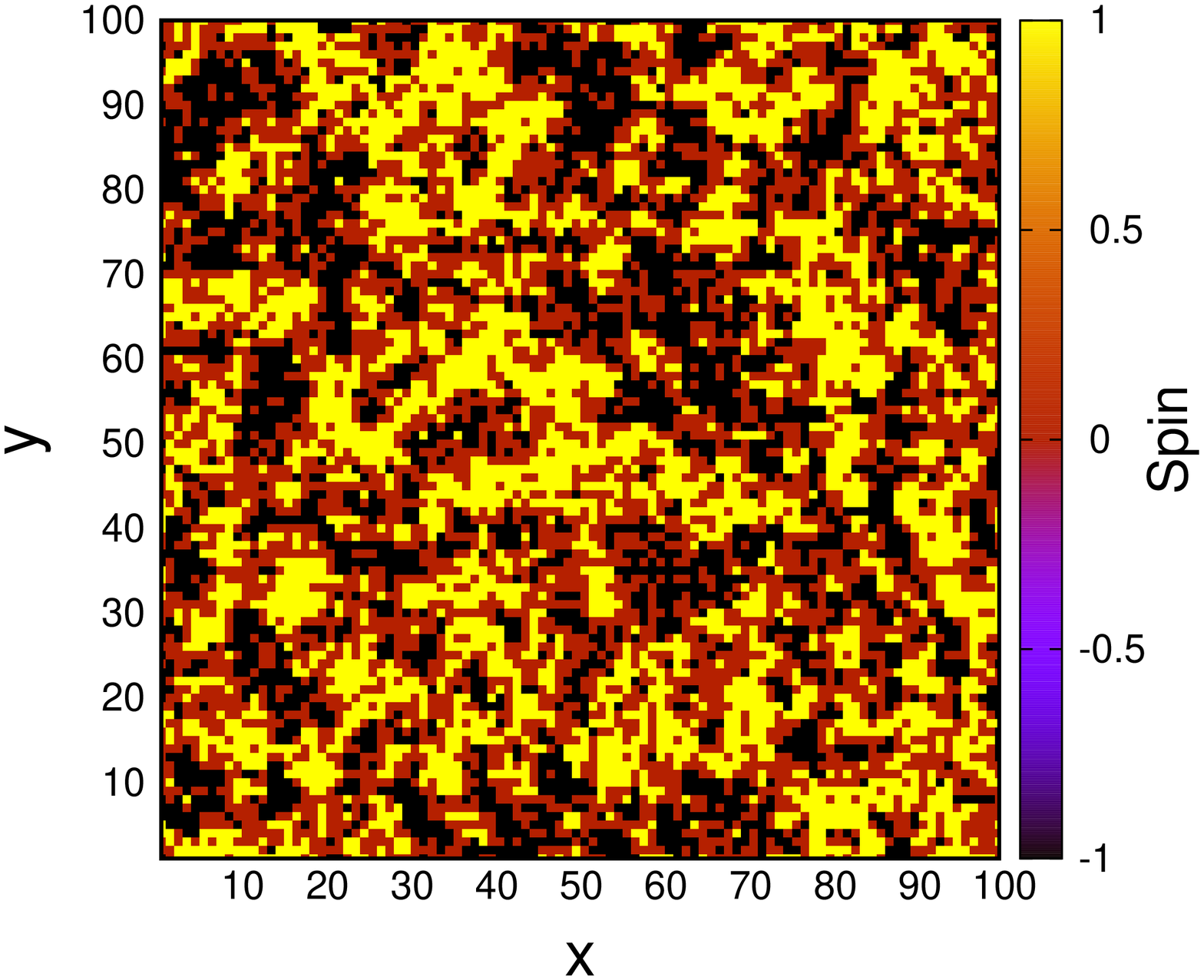}
	\subcaption{}
	\end{subfigure}
	\begin{subfigure}{0.333\textwidth}
	\includegraphics[angle=-90,width=\textwidth]{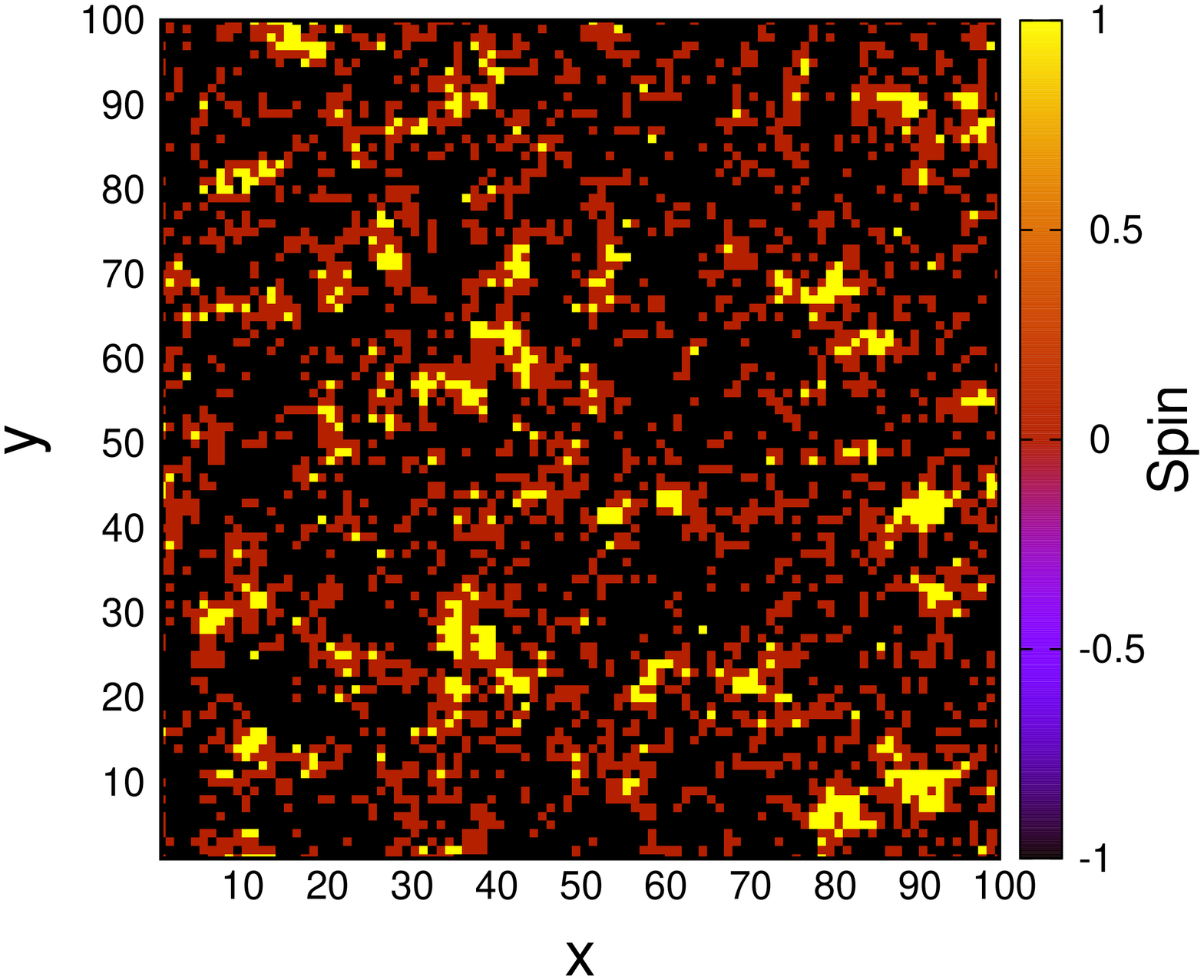}
	\subcaption{}
	\end{subfigure}
	\begin{subfigure}{0.333\textwidth}
	\includegraphics[angle=-90,width=\textwidth]{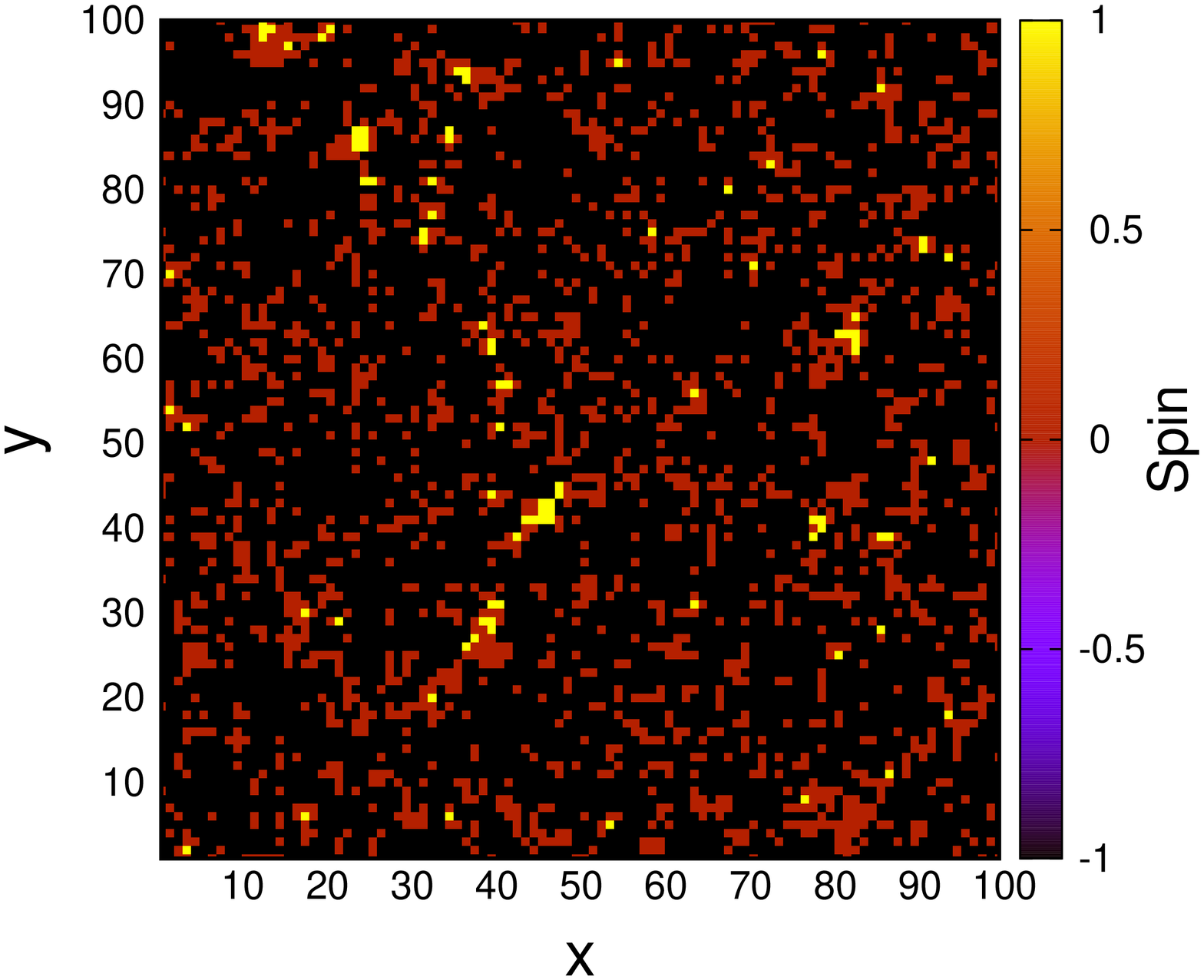}
	\subcaption{}
	\end{subfigure}
\caption{Snapshots of spin configurations captured at different times 
for the strength of anisotropy $D= 1.5$. (a) $t= 1$ MCSS (b) $t= 10$ MCSS 
(c) $t= 20$ MCSS (d) $t= 28$ MCSS(reversal time) (e) $t= 50$ MCSS
(f) $t= 70$ MCSS. Applied field is $h= -0.25$ and $T=1.2$.}
\label{snaps2}
\end{figure}

\begin{figure}[h!]
\begin{center}
\includegraphics[angle=-90,width=0.5\textwidth]{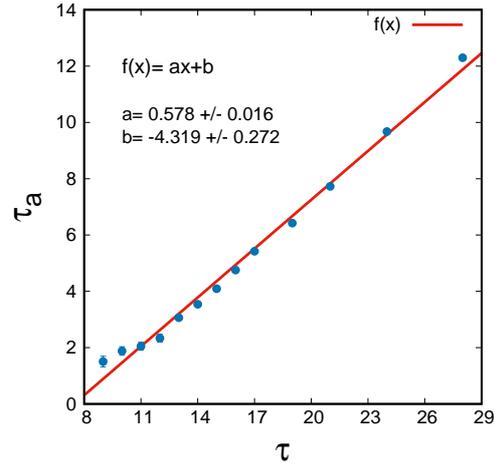}
\caption{Relation between microscopic reversal time ($\tau_a$) and
average macroscopic reversal time $\tau$. 
The value of $\tau_a = \frac{10}{b}$ is calculated from 
fig-\ref{revtime_confirm1}. Data are fitted to a straight line 
$f(x)= ax+b$ where $f(x)= \tau_a$ and $x= \tau$.}
\label{revtime_confirm2}
\end{center}
\end{figure}

{\it How does the macroscopic mean reversal time is connected to any microscopic scale of time ?}	
To address this interesting question,  we have 
studied the evolution of (in fig-\ref{revtime_confirm1}) density ($\rho_0^1$) of $S_i^z=0$, 
surrounded by all (four nearest neighbours) $S_i^z=+1$,  with time in presence of the anisotropy ($D>0$ here). 
That density $\rho_0^1$ is found to decay exponentially with time i.e. $\rho_0^1= a e^{-bt}$. The microscopic scale of time is defined from the exponential decay. 
So, $\frac{1}{b}= \tau_a$ is the microscopic scale of time in the present issue. 
Some snapshots (fig-\ref{snaps2})are 
captured at different time steps in presence of a particular anisotropy to 
have an idea about the evolution of the density $\rho_0^1$. Now the $\tau_a$ is 
determined for 
several values of anisotropy and plotted them with the reversal time ($\tau$) that 
we have defined earlier. It follows a straight line (fig-\ref{revtime_confirm2},
$\tau_a \sim c \tau$, where $c$ is a constant). 
This interesting observation prompted us
to have the idea of getting microscopic scale of time
($\tau_a$) which is related to the
macroscopic reversal time ($\tau$). It may be noted here, that both time scales are
measured in the case of single sample only (no averaging is carried out over different
random sample).

\begin{figure}[h!]
	\begin{subfigure}[b]{0.5\textwidth}
	\includegraphics[angle=-90,width=\textwidth]{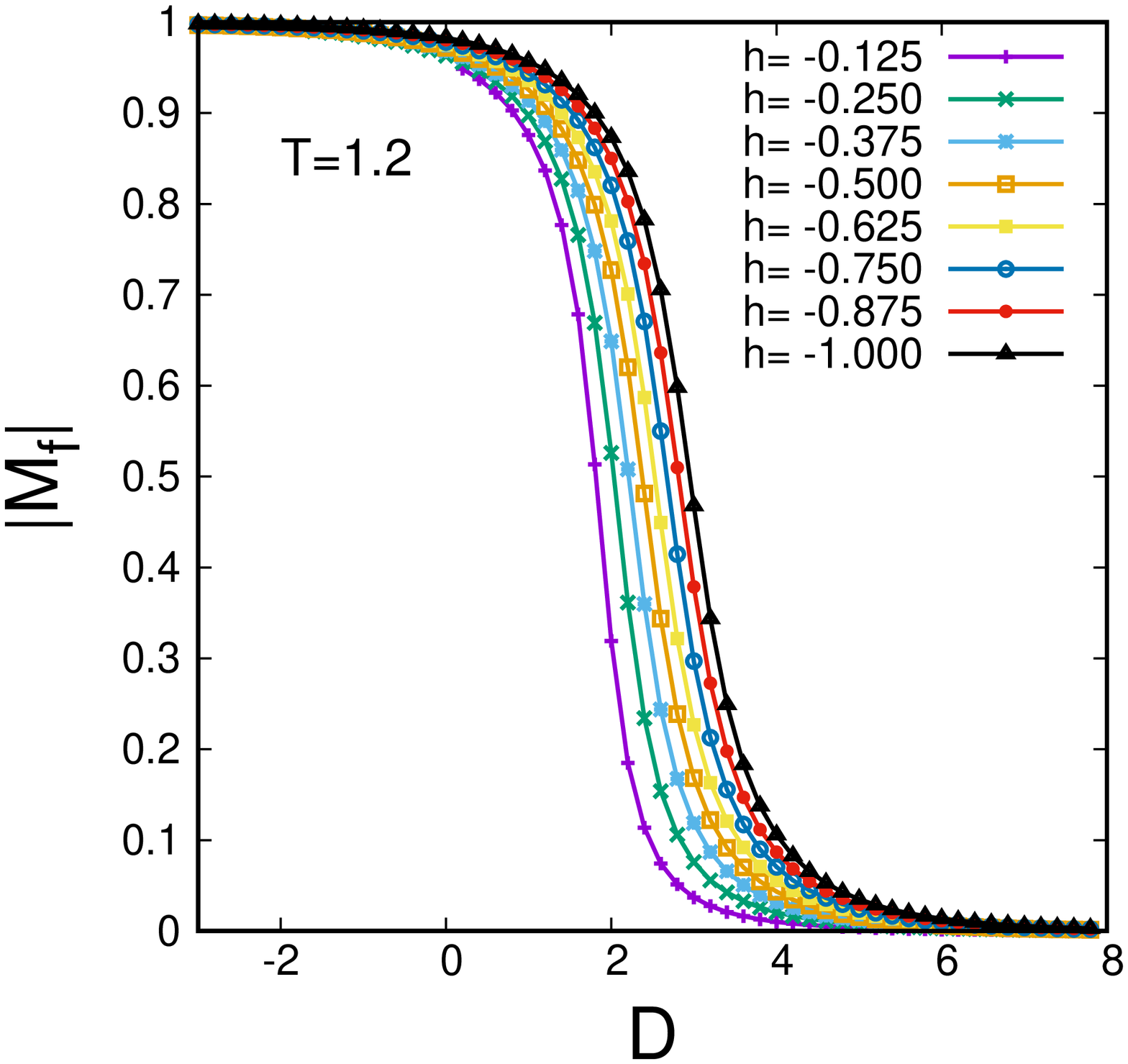}
	\subcaption{}
	\end{subfigure}
	\begin{subfigure}[b]{0.5\textwidth}
	\includegraphics[angle=-90,width=\textwidth]{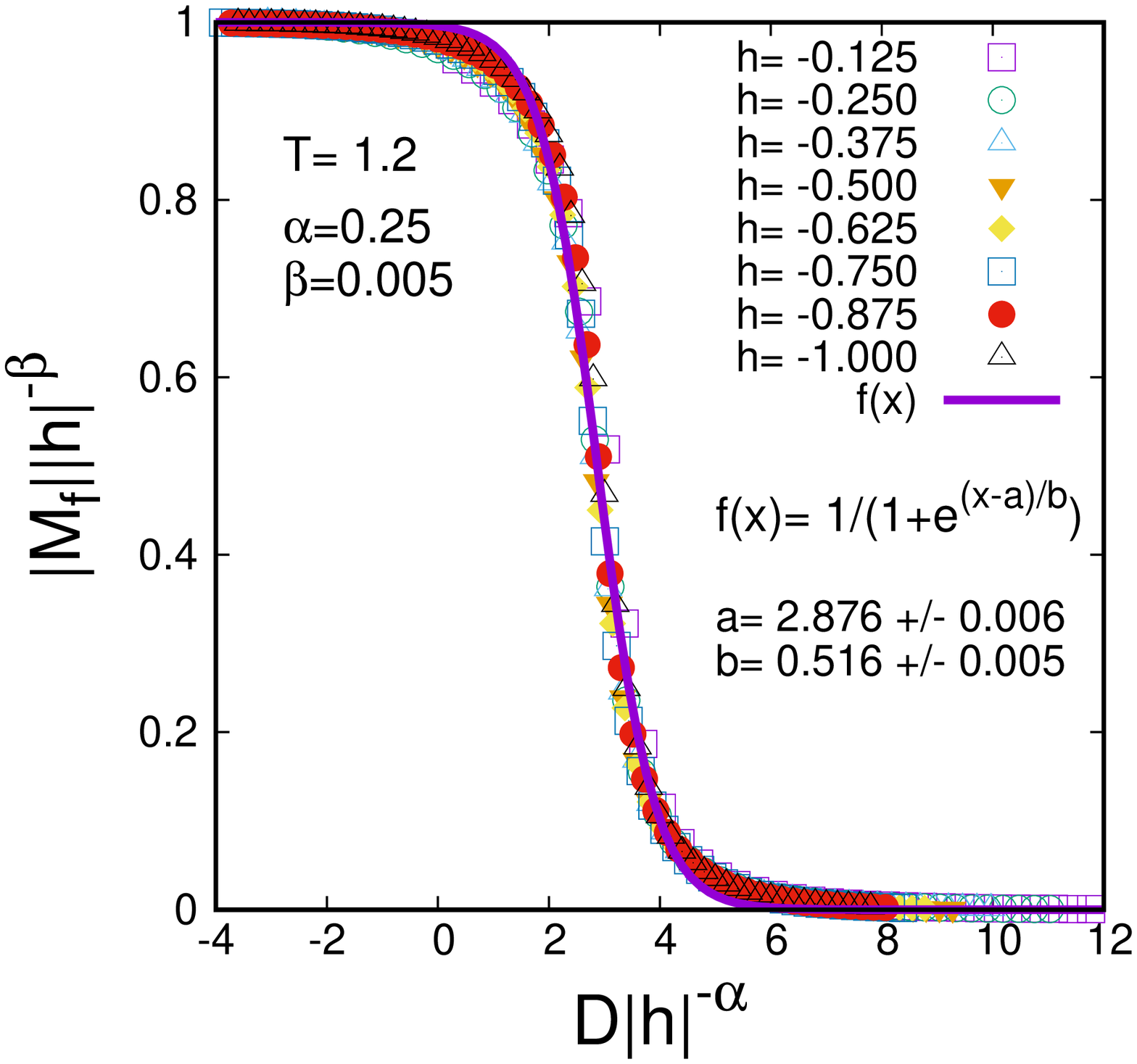}
	\subcaption{}
	\end{subfigure}
\caption{(a) Variation of saturated magnetisation ($M_f$) with 
anisotropy $(D)$ at any fixed temperature $T=1.2$ for different values of 
applied field ($h$), (b) Scaled saturated magnetisation ($|M_f||h|^{-\beta}$) 
versus scaled anisotropy ($D|h|^{-\alpha}$) at fixed temperature $T=1.2$.}
\label{scaling1}
\end{figure}

In fig-\ref{magtime} we observed that the anisotropy ($D$) of the system affects 
the saturation magnetisation after complete reversal ($|M_f|$). Now in 
fig-\ref{scaling1}a, the dependence of ($|M_f|$) on the anisotropy of the system 
is checked in presence of different values of applied field at a fixed temperature. 
Now all the plots are scaled (Fig-\ref{scaling1}b) by rescaling the anisotropy 
from $D$ to $D_s= D|h|^{-\alpha}$ and the saturation magnetisation from $|M_f|$ 
to $(|M_f|)_s$= $|M_f||h|^{-\beta}$. All the data points are observed to be collapsed 
to a single plot which is now fitted (Fig-\ref{scaling1}) to a scaling
function, $f(x)= \frac{1}{1+e^{(x-a)/b}}$ where $f(x)= |M_f||h|^{-\beta}$ and 
$x=D|h|^{-\alpha}$. From fig-\ref{scaling1}a, it is observed that scaling 
exponent $\alpha$ plays the crucial role ($\beta$ is quite small)
 in collapsing the data. 
The $|M_f|$ changes very slowly in the 
region of negative anisotropy ($D<0$) and also in the region of
 strong positive anisotropy
($D>0$) of the 
system. Because, in the presence of negative anisotropy ($D<0$), all the spins try to align 
along (parallel or antiparallel) the direction of 
the applied field. In contrast, the strong positive anisotropy
($D>0$) forbids the spins to be 
aligned along (parallel or antiparallel) the direction of 
 the applied magnetic field.
In between these two regions $|M_f|$ changes rapidly because, as 
the positive anisotropy becomes stronger, probability of microscopic transition 
from $S_i^z=+1$ to $S_i^z=0$  
 increases. So the density of $S_i^z=0$ increases with the increase 
in anisotropy causing significant reduction in $|M_f|$.
	
\begin{figure}[h!]
	\begin{subfigure}[b]{0.5\textwidth}
	\includegraphics[angle=-90,width=\textwidth]{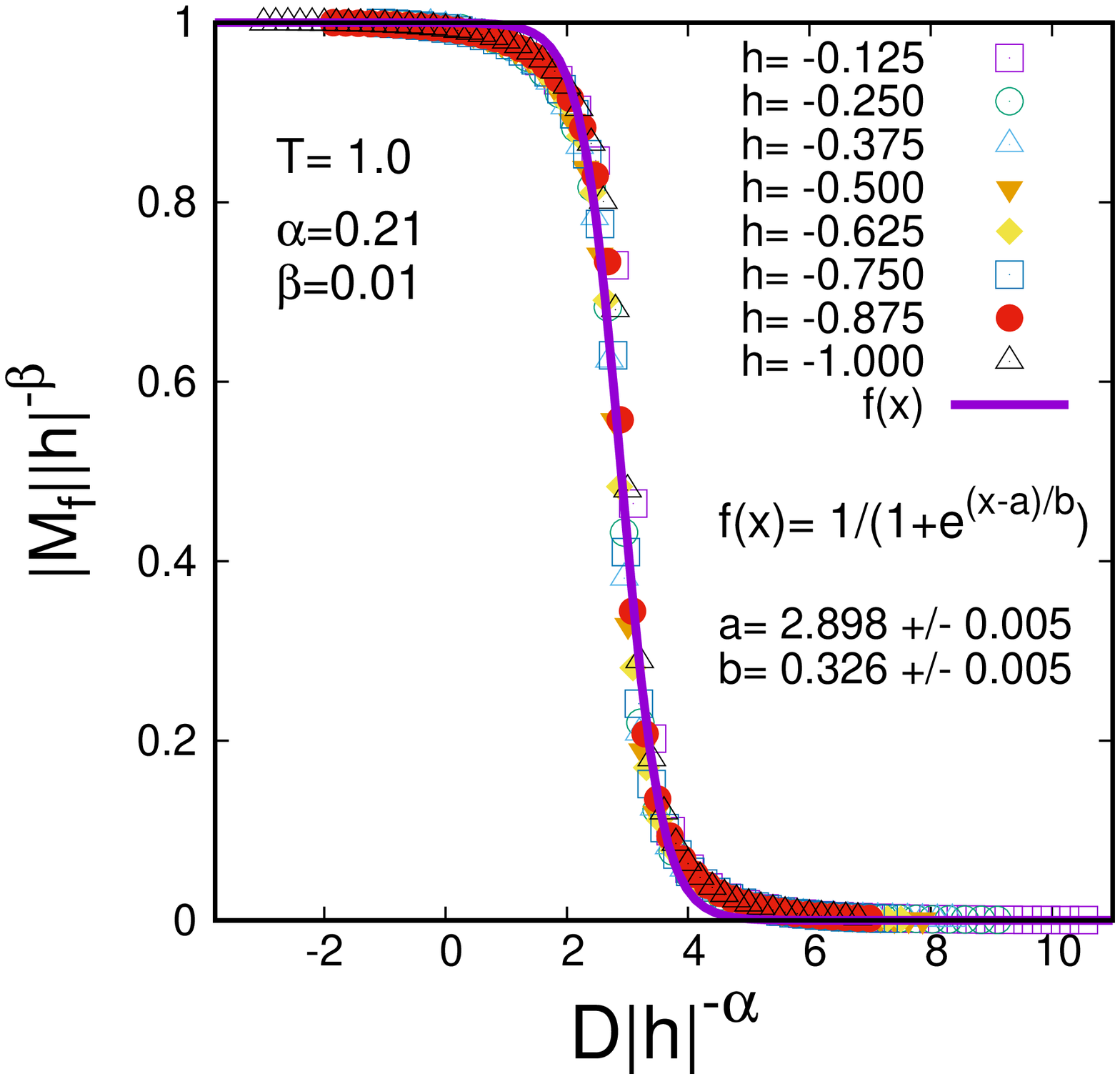}
	\subcaption{}
	\end{subfigure}
	\begin{subfigure}[b]{0.5\textwidth}
	\includegraphics[angle=-90,width=\textwidth]{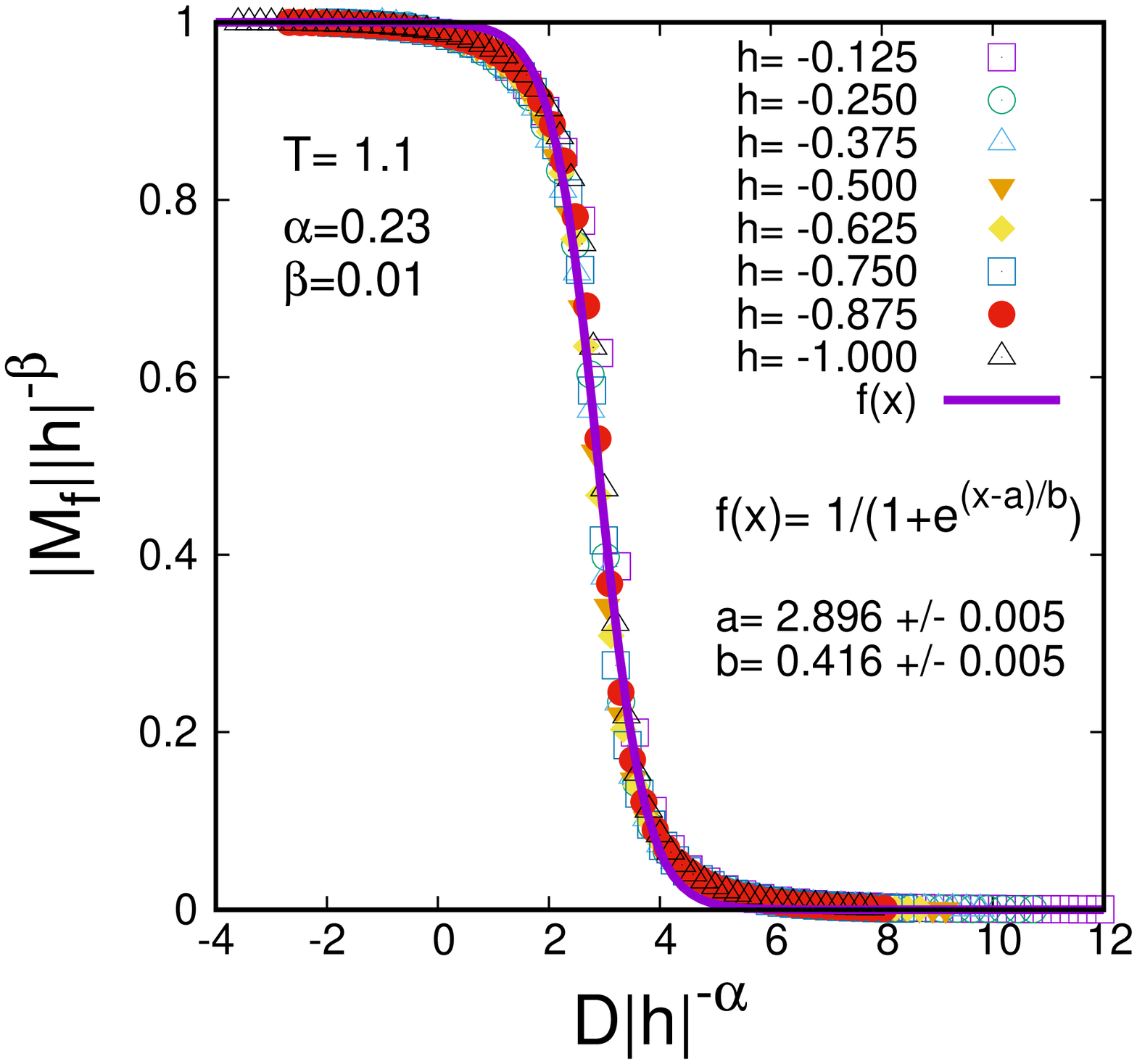}
	\subcaption{}
	\end{subfigure}
	\begin{subfigure}[b]{0.5\textwidth}
	\includegraphics[angle=-90,width=\textwidth]{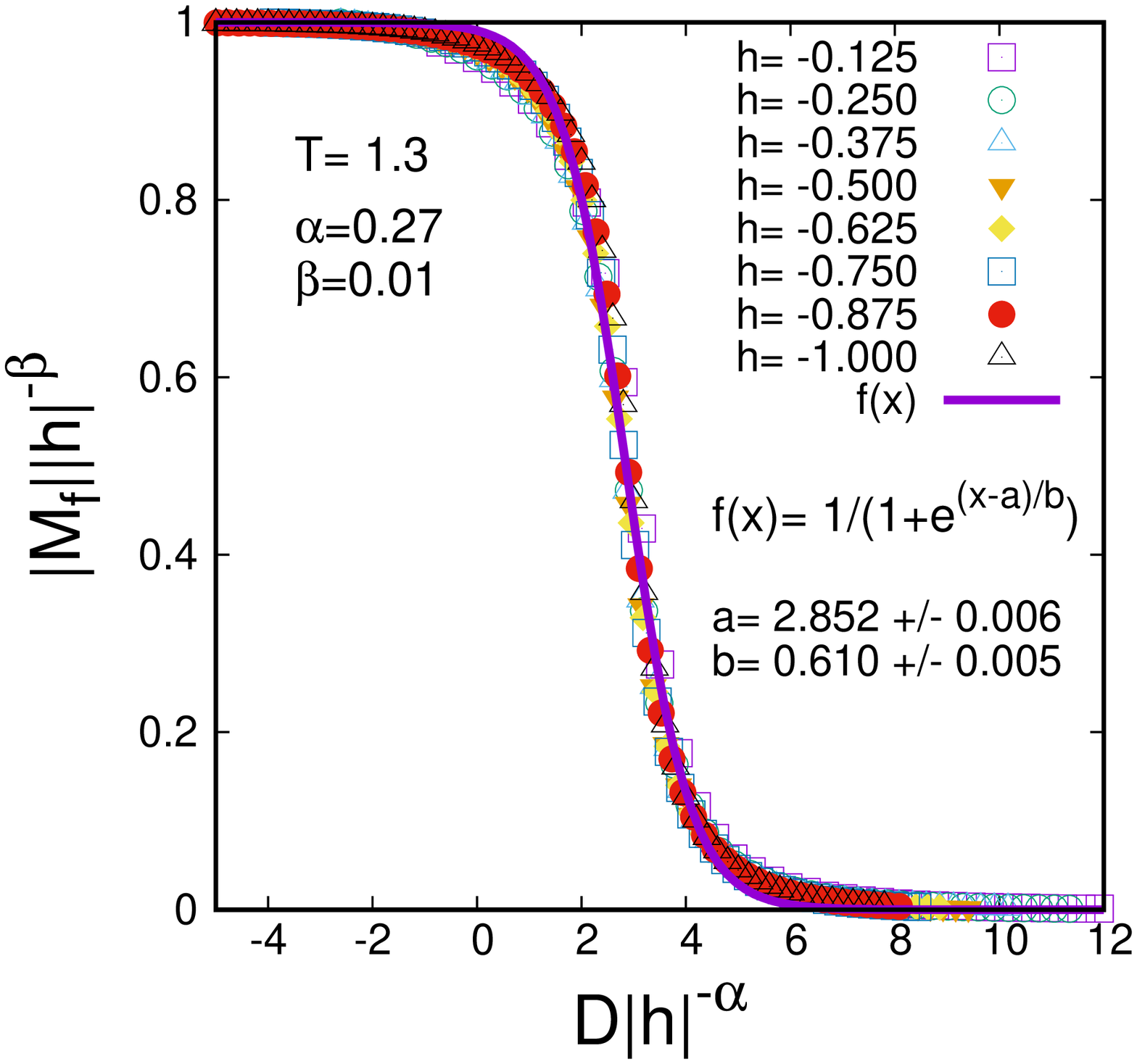}
	\subcaption{}
	\end{subfigure}
	\begin{subfigure}[b]{0.5\textwidth}
	\includegraphics[angle=-90,width=\textwidth]{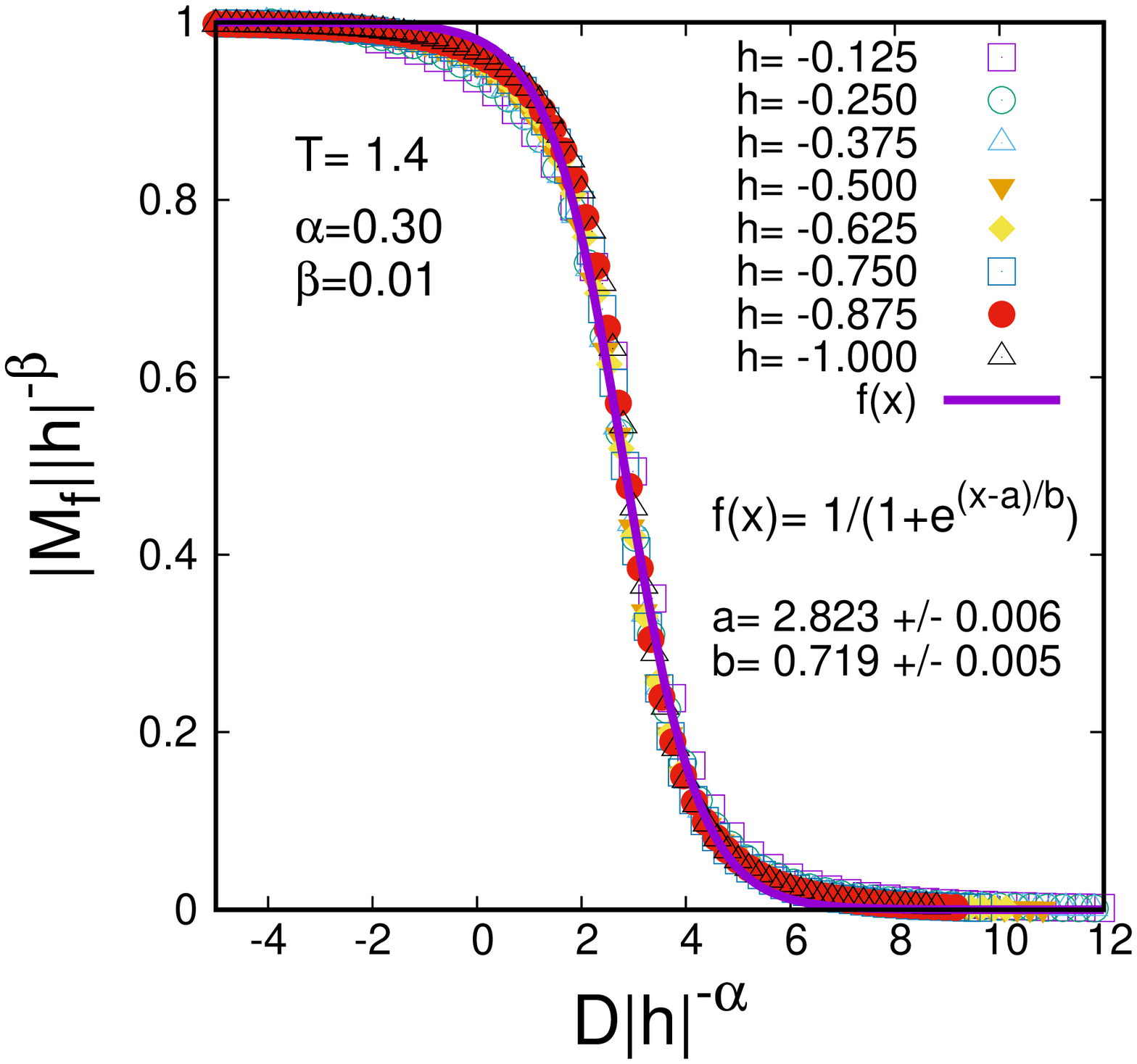}
	\subcaption{}
	\end{subfigure}
\caption{Scaling behaviour for another five different values of temperatures 
(a) $T= 1.0$ (b) $T= 1.1$(c) $T= 1.3$(d) $T= 1.4$}.
\label{scaling_temp}
\end{figure}

\begin{figure}[h]
\centering
\begin{subfigure}[b]{0.45\textwidth}
\includegraphics[angle=-90,width=\textwidth]{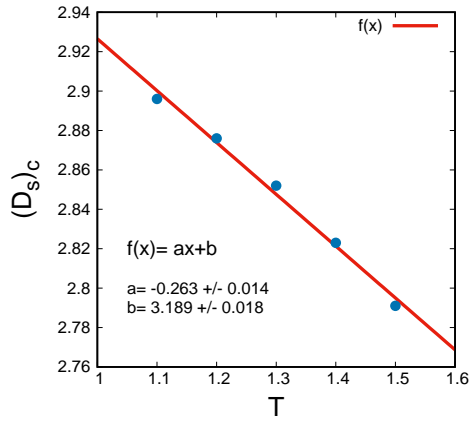}
\subcaption{}
\end{subfigure}
\begin{subfigure}[b]{0.45\textwidth}
\includegraphics[angle=-90,width=\textwidth]{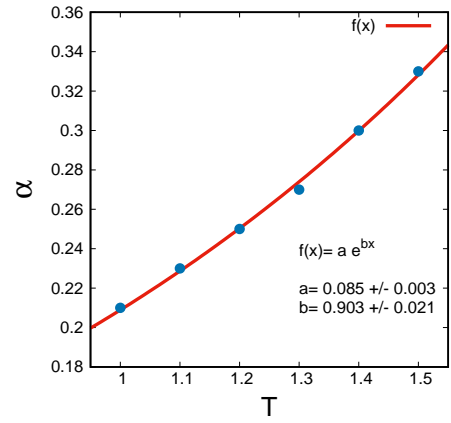}
\subcaption{}
\end{subfigure}
\caption{(a) Variation of the $(D_s)_c$ with temperature $T$. The data
are fitted with a straight line and (b) the variation of the 
scaling exponent $\alpha$ with temperature 
$(T)$. The data are fitted to the function $f(x)= a e^{bx}$ where 
$f(x)= \alpha$ and $x= T$.}
\label{scaling_expo}

\end{figure}

Above scaling analysis has been studied for different temperatures 
(fig-\ref{scaling_temp}). Below the temperature $T=1.0$ and above $T=1.5$ 
this scaling behaviour was not found to show significantly good data collapse. 
So, for a certain range (approximately 
from $T=1.0$ to $T=1.5$) of temperature, the system shows a fair scaling behaviour (as
verified by good data collapse). 
Within this range, for six different temperatures, we have obtained the values of 
the scaling exponent $\alpha$ by simple trial and error method. And also fitted 
(fig-\ref{scaling_temp}) to the function $f(x)= \frac{1}{1+e^{(x-a)/b}}$ 
similarly as fig-\ref{scaling_temp}. The value of $D_s$ at which ($(D_s)_c$ 
is actually the value of the fitting parameter $a$) $(|M_f|)_s$ becomes 0.5, is 
determined for each case. It ($(D_s)_c$) was found to decrease 
linearly (fig-\ref{scaling_expo}a) as the temperature
($T$) is increased. On the other hand, the scaling 
exponent $\alpha$ is observed to increase exponentially with the increase in 
temperature (fig-\ref{scaling_expo}b). Since the value of other scaling exponent
$\beta$ is too small to study any systematic variation with temperature.
 
\begin{figure}[h!]
	\begin{center}
		\includegraphics[angle=-90,width=0.5\textwidth]{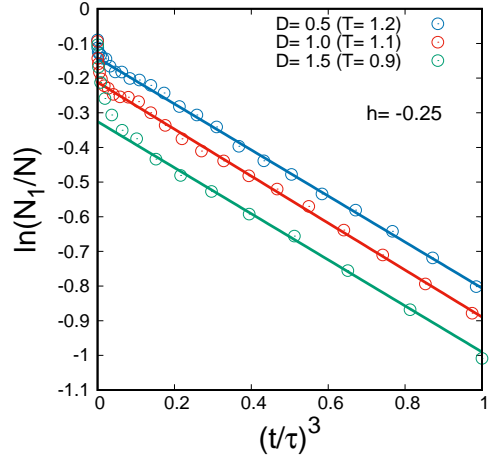}
		\caption{Evolution of metastable volume fraction (ln($\frac{N_1}{N}$), 
			where $N_1$ is the number of $S_i^z=+1$ and $N$ is the total number of 
			spins in the lattice) with cube of time ($(\frac{t}{\tau})^3$,
			 where $\tau$ is the reversal time) at three different values 
			 of positive anisotropy in presence of applied field $h= -0.25$. 
			 Temperature is fixed at $T=0.8 T_c$ for each case ($T=1.2$ and 
			 $\tau= 201$ for $D=0.5$, $T=1.1$ and $\tau=116$ for $D=1.0$, $T=0.9$ 
			 and $\tau=75$ for $D=1.5$).  Data are fitted to the straight lines 
			 $f(x)= ax+b$ with (i) $a= - 0.664 \pm 0.009$, $b= -0.142 \pm 0.005$ 
			 for $D= 0.5$ (ii) $a= - 0.677 \pm 0.011$, $b= -0.212\pm 0.006$ for 
			 $D= 1.0$ (iii) $a= - 0.664 \pm 0.016$, $b= -0.325\pm 0.007$ for 
			 $D= 1.5$.}
		\label{avrami}
	\end{center}
\end{figure}

In the present study, we have also investigated whether the system obeys Avrami's law 
regarding the  decay of metastable
volume fraction in presence of anisotropy. According to the KJMA theory, metastable 
volume fraction decays exponentially with a power of time. Avrami's law depicts that metastable 
volume fraction (relative abundance of $S_i^z=1$) 
decays exponentially with $t^{d+1}$ in 
a $d$ dimensional system (closer to the critical temperature $T= 0.8 T_c$ here). 
So for a 
two dimensional Blume-Capel ferromagnet, the logarithm of the metastable volume fraction is plotted 
(fig-\ref{avrami}) against the third power of time i.e., ($\frac{t}{\tau})^3$ ( nondimensionalised 
by reversal time $\tau$) which fits fairly to a straight line. This confirms that
the Avrami's law holds good in the case of $S=1$ Blume-Capel ferromagnetic system
with anisotropy. We have 
checked that law in the presence of three different strength of anisotropy ($D= 0.5$, 
$D= 1.0$ and $D= 1.5$) (shown in Fig-\ref{avrami})
keeping the temperature of the system fixed at $T= 0.8 T_c$ for 
each case \cite{butera}. Here also, the system is found to 
follow the Avrami's law for different values of $D$.

\begin{figure}[h!]
	\begin{subfigure}{0.5\textwidth}
	\includegraphics[angle=-90,width=\textwidth]{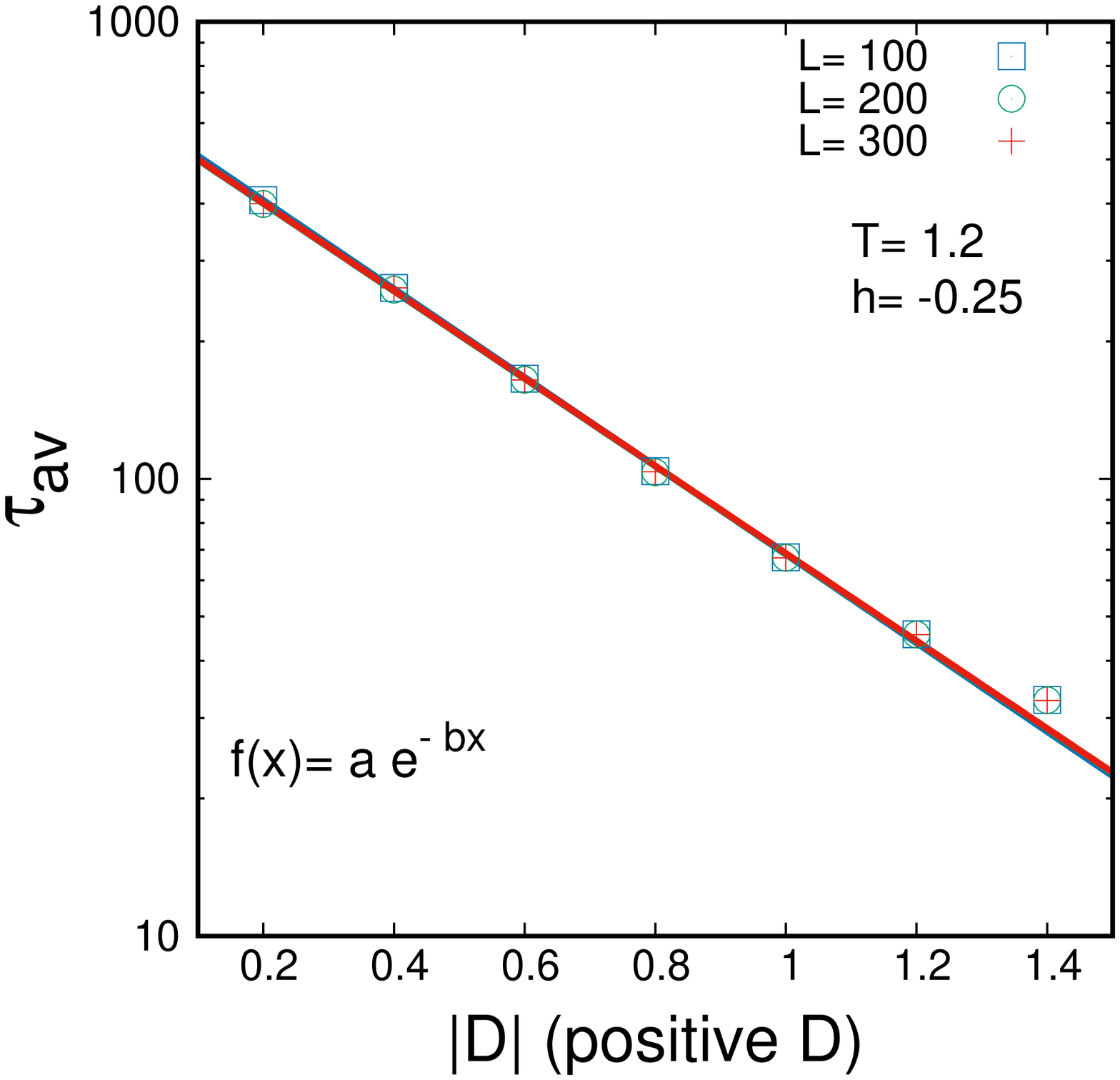}
	\subcaption{}
	\end{subfigure}
	\begin{subfigure}{0.5\textwidth}
	\includegraphics[angle=-90,width=\textwidth]{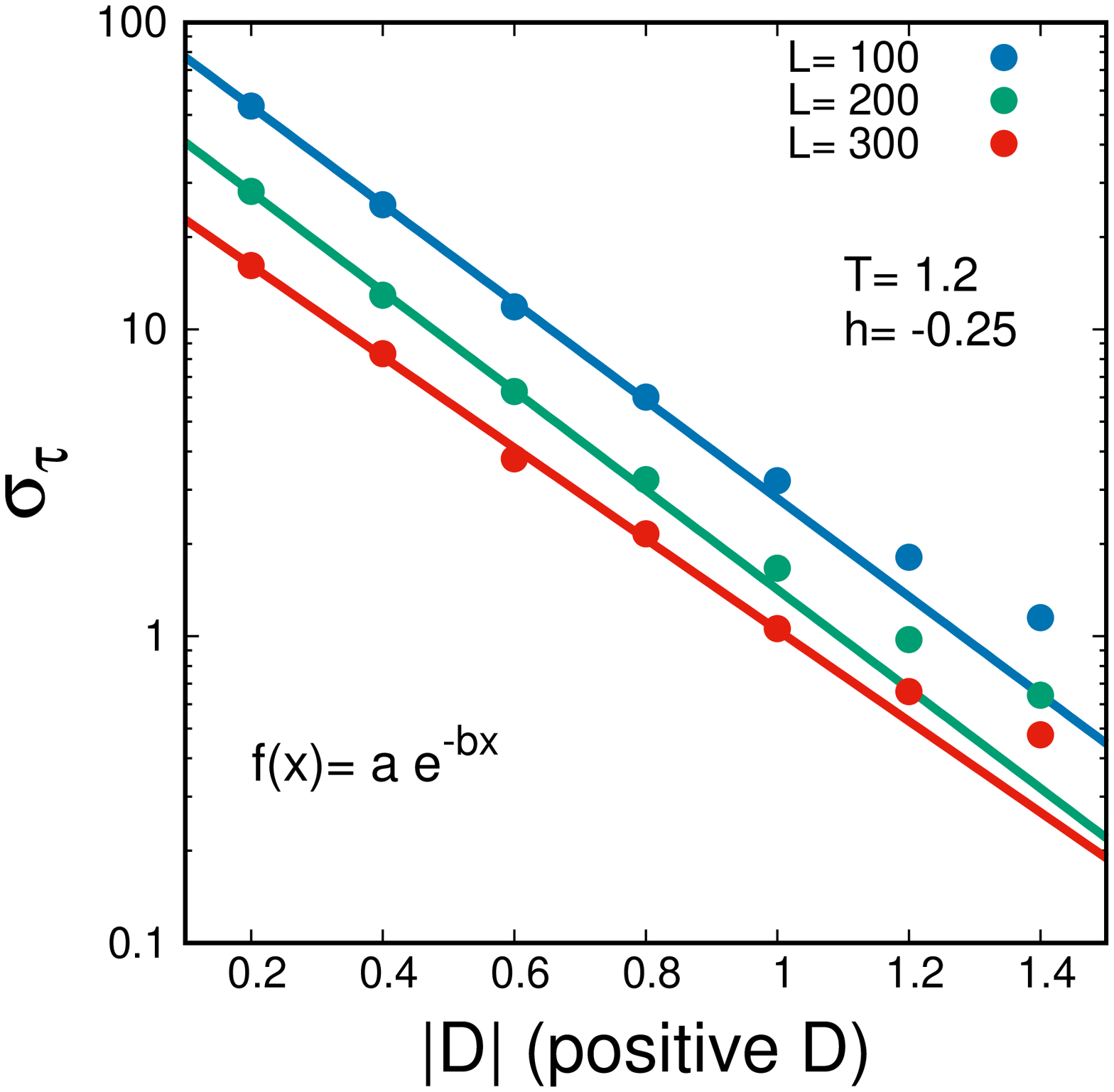}
	\subcaption{}
	\end{subfigure}
	
	\caption{ (a) Semilogarithmic plot of mean reversal time against 
			positive anisotropy for different size of lattice ($L= 100$, 
			$L= 200$ and $L= 300$) at fixed temperature $T=1.2$ in 
			presence of applied field $h= -0.25$. Data are fitted to the 
			function $f(x)= a e^{-bx}$ where $x= |D|$ and $f(x)= \tau_{av}$ 
			with (i) $b= 2.23 \pm 0.02$ for $L= 100$ (ii) $b= 2.205 \pm 0.026$ 
			for $L= 200$ (iii) $b= 2.205 \pm 0.028$ for $L= 300$. 
			(b)  Semilogarithmic plot of standard deviation of mean reversal 
			time against positive anisotropy for different size of lattice ($L= 100$, 
			$L= 200$ and $L= 300$) at fixed temperature $T=1.2$ in presence of 
			applied field $h= -0.25$. Data are fitted to the function 
			$f(x)= a e^{-bx}$ where $x= |D|$ and $f(x)= \sigma_\tau$ with 
			(i) $b= 3.68 \pm 0.05$ for $L= 100$ (ii) $b= 3.73 \pm 0.08$ 
			for $L= 200$ (iii) $b= 3.42 \pm 0.08$ for $L= 300$. }
	\label{finitesize}
\end{figure}

The cooperatively interacting many body system, with large degrees of freedom,
is found to suffer from the finite size 
effect. To study such effect, we have investigated the  mean reversal time (obtained 
from 1000 samples) and also its standard deviation against strength of anisotropy for 
three different size of lattice
($L=100, 200$ and 300). Fig-\ref{finitesize} depicts that the mean reversal time 
$\tau_{av}$ is beyond any finite size effect in the system size considered in our
present study. On the other hand, the standard deviation 
$\sigma_{\tau}$ of reversal times, decreases for larger system sizes. This is 
obviously in conformity with general statistical mechanical study. 

\begin{figure}[htb]
	\begin{subfigure}{0.5\textwidth}
		\includegraphics[angle=-90,width=\textwidth]{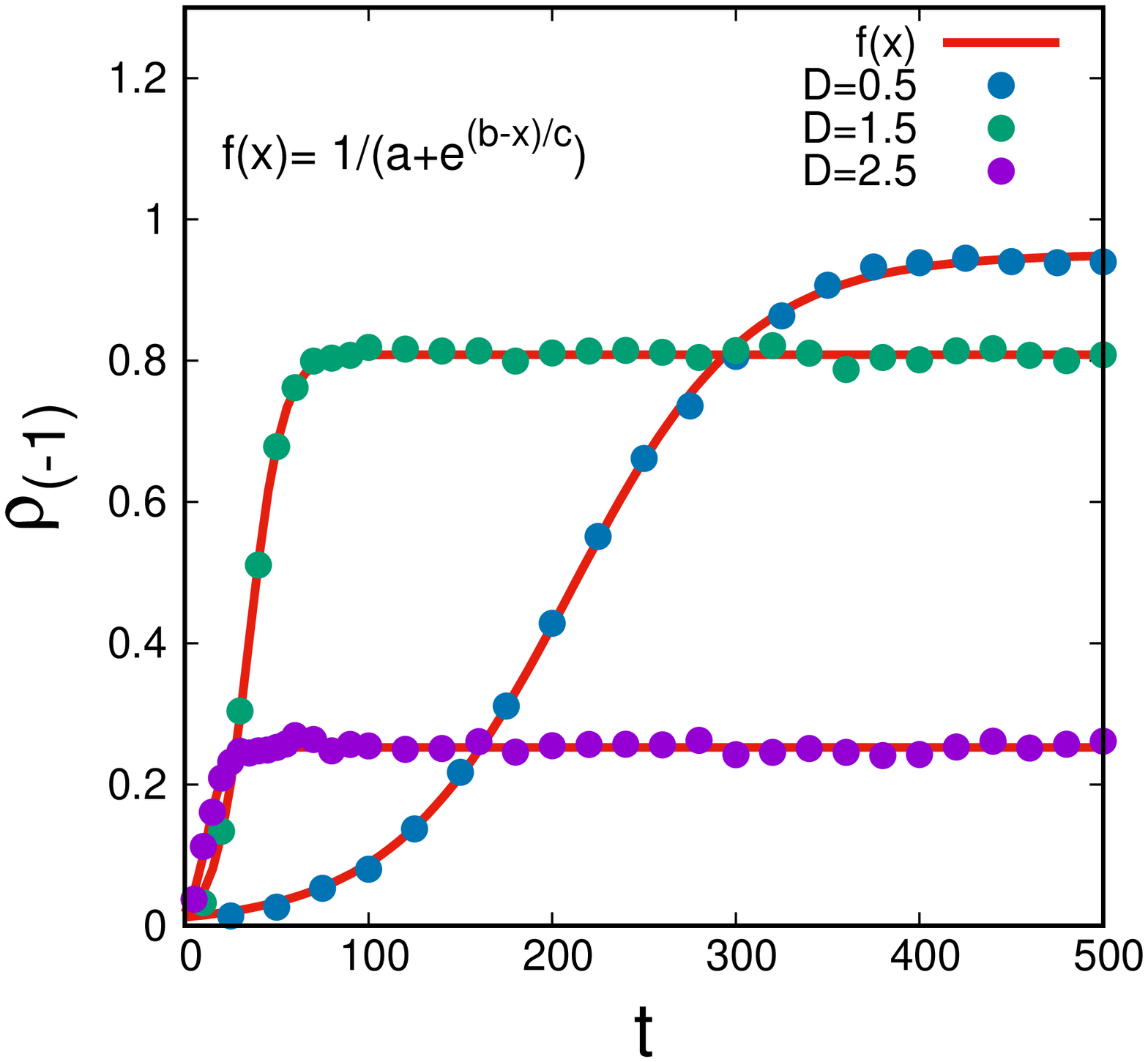}
		\subcaption{}
	\end{subfigure}
	\begin{subfigure}{0.5\textwidth}
		\includegraphics[angle=-90,width=\textwidth]{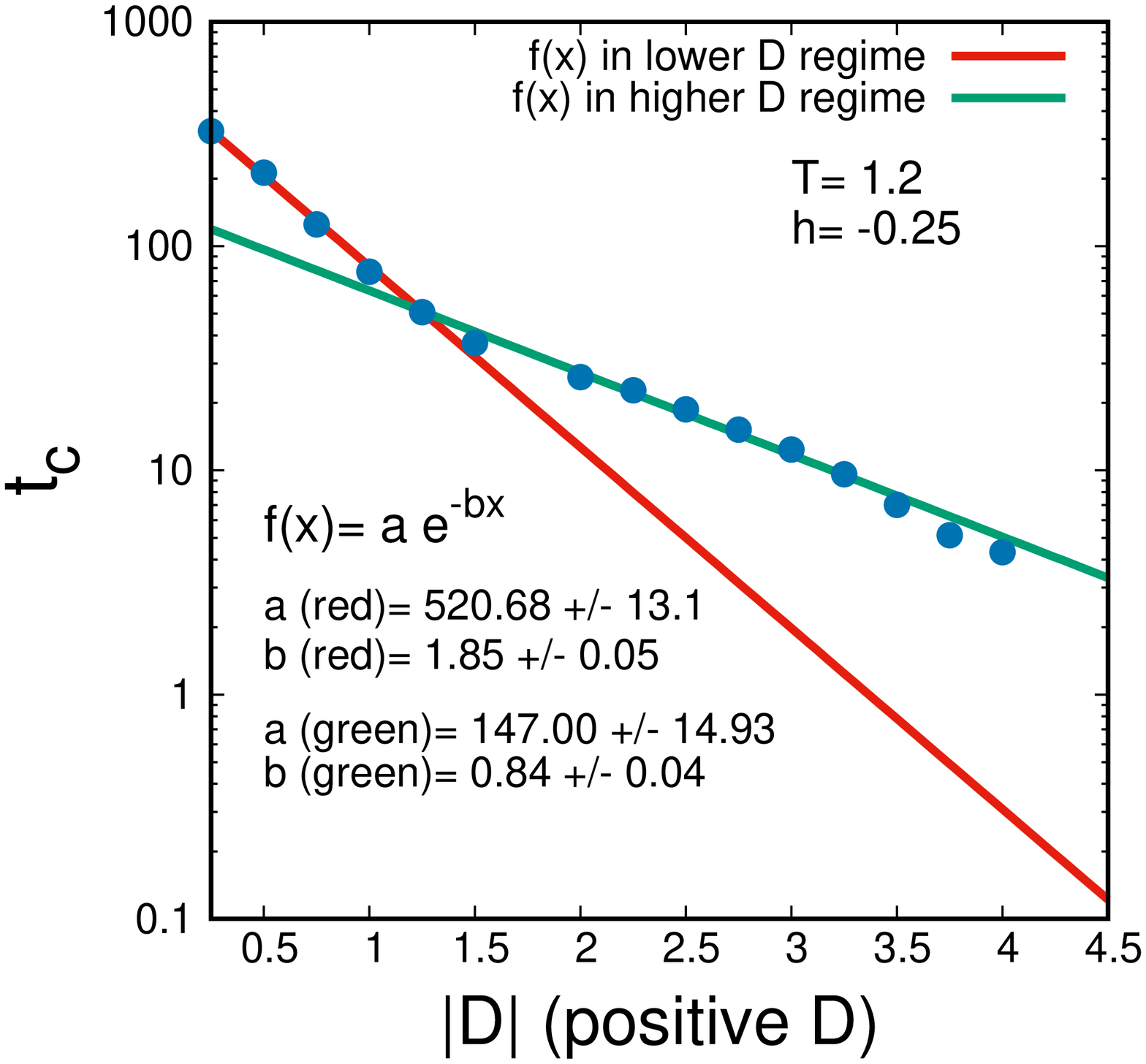}
		\subcaption{}	
	\end{subfigure}

	\caption{(a) Temporal evolution of density ($\rho_{(-1)}$) of $S_i^z=-1$  
		at three different values of magnetic anisotropy ($D= 0.5$, $D=1.5$,$D=2.5$) 
		which are fitted to the function $f(x)= \frac{1}{a+e^{(b-x)/c}}$ 
		where $f(x)= \rho_{(-1)}$ and $x=t$. Temperature is set to $T=1.2$ 
		and applied field is $h= -0.25$. Values of the parameters are 
		(i) $a= 1.052 \pm 0.001$, $b= 212.793 \pm 0.155$, $c= 48.435 \pm 0.144$ 
		for $D= 0.5$ (ii) $a= 1.236 \pm 0.001$, $b= 36.902 \pm 0.071$, 
		$c= 9.025 \pm 0.062$ for $D= 1.5$ (iii) $a= 3.96 \pm 0.005$, 
		$b= 18.70 \pm 0.255$, $c= 4.80 \pm 0.150$ for $D= 2.5$. (b) Variation 
		of characteristic time ($t_c$) or the values of parameter b (obtained 
		from (a)) with positive D. Data are fitted to the exponential function 
		separately in two regimes.}
	\label{denm1}
\end{figure}

Variation of the density of $S_i^z=-1$ ($\rho_{(-1)}$) is studied 
as function of time and shown in fig-\ref{denm1}a 
for three different values of anisotropy $D$. The density 
$\rho_{(-1)}$,
is found to fit with a function like
$f(x)= \frac{1}{a+e^{(b-x)/c}}$. Stronger anisotropy raises the parameter $a$, whereas 
reduces the the value of $b$. From the function $f(x)$, a characteristic time ($t_c$) (i.e. 
the value of the parameter $b$ in a sense) is defined at which $f(x)= \rho_{(-1)}= \frac{1}{a+1}$.
Whereas $\frac{1}{a}$ is the saturated density of $S_i^z=-1$, achieved after a very
long 
of time ($t \to \infty$). The $t_c$ can give an approximate idea about the 
influence of anisotropy on the rate of
growth of density of $S_i^z=-1$. The variation of $t_c$ with $D$ (positive) is also
studied here (fig-\ref{denm1}b).The data are exponentially fitted separately in two regimes of $D$. 
It is observed that $t_c \sim e^{-rD}$. Interestingly,
the value of $r=1.85$ in the weak anisotropy regime and $r=0.84$ in the strong
anisotropy regime. A crossover is observed around $D=1.5$. 
In the presence 
of stronger magnetic anisotropy $D$, the rate of achieving the density $\frac{1}{a+1}$ is faster than that in the regime of weaker $D$. Because, in stronger $D$ regime, density of $S_i^z=0$, plays a dominating role. The change from $S_i^z=+1$ to $S_i^z=-1$ in the low density of $S_i^z=0$, is less probable than that for high
density of $S_i^z=0$. As a result, the density of $S_i^z=-1$ takes shorter time to saturate in the limit of high $D$. 
%----------------------------------------------------------------------------------
\vskip 2 cm
\noindent {\large\bf IV. Summary} 

\vskip 0.2 cm

In this article, the reversal of magnetisation is studied in a two dimensional
anisotropic ($S=1$) Blume-Capel ferromagnet by Monte Carlo simulation using
Metropolis single spin flip algorithm. The reversal time was studied in details
as function of the strength of the anisotropy $D$. The statistical distribution of the
reversal times (for different random samples) was also studied. The most probable
value of this distribution was found to be strongly dependent of $D$. The most
probable reversal time decreases as the strength of anisotropy $D (>0)$ increases.
The mean (or average) reversal time is found to be exponentially decreasing with
$D$. On the other hand, the mean reversal time was found to increase exponentially
with the magnitude of $D$, for negative anisotropy. The standard deviation of reversal
times shows an interesting behaviour. For positive $D$, it falls exponentially with
the magnitude of $D$. However, the rate of falling is higher in the low anisotropy
regime. In the case of negative anisotropy, a qualitatively contrast behaviour is
observed. We have also studied the evolution of densities of the different values
of $S_i^z$ i.e. +1, 0 and -1 in a microscopic level for better understanding of
 the reversal
phenomena in the Blume-Capel model. The mean density $\rho_{0}^1$, i.e. density of
$S_i^z=0$ (surrounded by all $S_i^z=+1$) is studied as function of time. It is observed
to decay exponentially with time. Here, the microscopic time is found to depend 
linearly on the macroscopic reversal time.

After the reversal, the magnetisation was observed to reach
 a saturated value (with some fluctuation
of course). This saturated magnetisation $M_f$, after the reversal, is found to be
strongly
dependent on the strength of anisotropy $D$. An interesting scaling relation was found $|M_f| \sim |h|^{\beta}f(D|h|^{-\alpha})$ with the 
scaling function of the form $f(x)= \frac{1}{1+e^{(x-a)/b}}$. The scaling relation was obtained in a certain range of temperature ($T$) by employing the method of simple data collapse. The scaling exponent $\alpha$ has a dependence on the temperature ($T$)
of the system.

 The growth of mean density of $S_i^z=-1$ is observed to be fitted in a
function like $\rho_{-1}(t) \sim \frac{1}{a+e^{(t_c-t)/c}}$. 
The characteristic
time ($t_c$) behaves like $t_c \sim e^{-rD}$. A crossover of the values of $r$ is 
detected at $D=1.5$. 

We have also studied the functional form of the metastable volume fraction
(i.e., relative abundance of $S_i^z=+1$). It is observed the the metastable
volume fraction decays exponentially with the third power of time ($t$) indicating
the Avrami's law.	

The size dependences of the reversal time and its fluctuations are also checked for three different
system sizes. The mean reversal time does not show any remarkable change however
the standard deviation of reversal time decreases with the increase of the system
size which is quite expected in general statistical analysis.

In the present study, we have considered both the positive and negative values of the anisotropy $D$. The two dimensional Blume-Capel model shows an interesting phase
diagram (of ferro-para phase transition) in the plane formed by crystal field and the
temperature. We have not found any significant change in the monotonic behaviour 
of the reversal time while acrossing the phase boundary. This has been reported in another
article\cite{gradient}. But we should keep in mind that the reversal is triggered 
under the application of nonzero external magnetic field, which is zero in the phase diagram
mentioned in this context.

The functional forms of the, saturated magnetisation after reversal depending on the anisotropy and field, are obtained from the numerical data. It would be interesting to derive those laws by analytic methods. For 
this reason the Becker-D\"oring analysis, for the growth of critical clusters, has to be extended for anisotropic Blume-Capel model. The analytic derivation of size dependences of the mean reversal time also demands such extension of classical nucleation theory.

%----------------------------------------------------------------------------------
\vskip 2 cm
\noindent {\large\bf V. Acknowledgements}
	
MA would like to acknowledge FRPDF research grant provided by Presidency University.
MN would like to acknowledge Swami Vivekananda fellowship for financial support.

\vskip 0.2 cm

%--------------------------------------------------------------------------

%------------------------------------------------------------------------------------
%****FIGURES****%
%------------------------------------------------------------------------------------
%%%% FIG-1
\newpage

%%%%%FIG-2
\newpage

%*****FIG-3
\newpage
	
%*****FIG-4
\newpage

%******FIG-5
\newpage

%***FIG-6
\newpage

%*********FIG-7
\newpage

%*******FIG-8
\newpage

%********FIG-9
\newpage

%****FIG-10
\newpage	

%*******FIG-11
\newpage
		
%********FIg-12
\newpage
				
%****FIG-13
\newpage

%******FIG-14
\newpage

%**************************************************************

%****FIG-15
\newpage

%********FIG-16
\newpage

%--------------------------------------------------------------------------------

\vskip 2 cm
\begin{thebibliography}{99}
\bibitem{techno} S. N. Piramanayagam and T. C. Chong, Development in
	data storage: Material perspective, Wiley-IEEE Press, 2011
\bibitem{daniel} E. D. Daniel, C. Denis Mes and M. H. Clark, Magnetic recording:
The first 100 years, Wiley-IEE Press, 1998
	\bibitem{becker} R. Becker and W. \"Doring, Ann. Phys. (Leipzig)
		416 (1935) 719
\bibitem{grant} M. Grant and J. D. Gunton, Phys. Rev. B 32 (1985) 7299

\bibitem{rikvold1} P.A. Rikvold, H. Tomita, 
	S. Miyashita, S.W. Sides, Phys. Rev. E 49 (1994) 5080.
\bibitem{stauffer} M. Acharyya, D. Stauffer, Eur. Phys. J. B. i
	5 (1998) 571.
\bibitem{bkc} A. Misra, B.K. Chakrabarti, Physica A 246 (1997) 510.
\bibitem{binder1} K. Binder and H. M\"uller-Krumbhaar, Phys. Rev. B
	9 (1974) 2328.
\bibitem{uli} D. Hinzke, U. Nowak, Phys. Rev. B. 58 (1998) 265.
\bibitem{vehkamaki} H. Vehkamäki, I.J. Ford, Phys. Rev. E 59 (1999) 6483.
\bibitem{kolmogorov} A.N. Kolmogorov, Bull. Acad. Sci. 
	USSR Ser. Math. 3 (1937) 355.
\bibitem{johnson} W.A. Johnson, P.A. Mehl, Trans. 
	Am. Inst. Min. Metall. Eng. 135 (1939) 416.
\bibitem{avrami} M. Avrami, J. Chem. Phys. 7 (1939) 1103; 8 (1940) 212; 9 (1941) 177.
\bibitem{rikvold2} W.R. Deskins, G. Brown, S.H. Thompson, P.A. Rikvold, Phys. Rev. B 84 (2011) 094431
\bibitem{ma1} M. Acharyya, Physica A 403 (2014) 94.
\bibitem{ma2} A. Dhar, M. Acharyya, Commun. Theor. Phys. 66 (2016) 563.
\bibitem{ma3} R. Dutta, M. Acharyya, A. Dhar, Heliyon 4 (2018) e00892.
\bibitem{moumita} M. Naskar and M. Acharyya, Physica A 551 (2020) 124583.
\bibitem{blume} M. Blume, Phys. Rev. 141 (1966) 517.
\bibitem{capel} H. Capel, Physica. 32 (1966) 966.
\bibitem{cirillo} E. N. M. Cirillo and E. Olivieri, J. Stat. Phys. 83
	(1996) 473
\bibitem{costabile} E. Costabile, M. A. Amazonas, J. R. Viana, 
	J. R. de Sousa, Phys. Lett. A 376 (2012) 2922
\bibitem{silva} C. J. Silva, A. A. Caparica, J. A. Plascak,
	Phys. Rev. E 73 (2006) 036702
\bibitem{gulpinar} G. Gulpinar, E. Vatansever, M. Agartioglu,
	Physica A 391 (2012) 3574.
\bibitem{ajay} M. Acharyya and A. Halder, J. Magn. Magn. Mater. 426
	(2017) 53.
\bibitem{park}Y. Yamamoto and K. Park, Phys. Rev. E 88 (2013) 012110. 		
\bibitem{fisher} J. M. Yeomans and M. E. Fisher, Phys. Rev. B, 24 (1981) 2825
\bibitem{kwak}W. Kwak, J. Jeong, J. Lee 
	and Dong-Hee Kim, Phys. Rev. E 92 (2015) 022134

\bibitem{fytas1} J. Zierenberg, N. G. Fytas, M. Weigel, W. Janke, A. Malakis
European Physical Journal Special Topics 226 (2017) 789
\bibitem{fytas2} N. G. Fytas, J. Zierenberg, P. E. Theodorakis, M. Weigel,
W. Janke, A. Malakis, Phys. Rev. E, 97 (2018) 040102(R)
\bibitem{erol2} E. Vatansever, Z. D. Vatansever, P. E. Theodorakis and N. G. Fytas,
Phys. Rev. E 102 (2020) 062138
\bibitem{sumedha} Sumedha and N. K. Jana, J. Phys A: Math. Theor. 50 (2017) 015003
\bibitem{selke} W. Selke and J. Oitmaa, J. Phys.: Condens. Matter 22 (2010) 076004
\bibitem{bahmad} R. Masrour, A. Jabar, L. Bahmad, M. Hamedoun, A. Benyoussef,
J. Magn. Magn. Mater. 421 (2017) 76

\bibitem{binder2} K. Binder and D.W Heermann, Monte Carlo Simulation in Statistical physics, Second edition,
Springer-Verlag (1992), Berlin
\bibitem{metro}N. Metropolis, A. W. Rosenbluth, M. N. Rosenbluth, A. H. Teller, The Journal of Chemical
Physics, Vol 21, Number 6, June 1953.
\bibitem{butera}P. Butera, M. Pernici, Physica A 507 (2018) 22.
\bibitem{gradient} M. Naskar and M. Acharyya, (2020), arxiv:2009.08342
\end{thebibliography}
\end{document}